\documentclass[12pt,preprint]{aastex}

\newcommand{\kev}{keV}
\newcommand{\etal}{et al.}
\newcommand{\nh}{$N_{\mathrm{H}}$}
\newcommand{\chandra}{\textit{Chandra}}
\newcommand{\xmm}{\textit{XMM-Newton}}

\shorttitle{Galaxy Evolution, Star Formation \& the CXRB}
\shortauthors{Ballantyne, Everett \& Murray}

\begin{document}

\title{Connecting Galaxy Evolution, Star Formation and the X-ray Background}


\author{D. R. Ballantyne\altaffilmark{1}, J. E. Everett and N. Murray}
\affil{Canadian Institute for Theoretical Astrophysics,
McLennan Labs, 60 St. George Street, Toronto, Ontario, Canada M5S 3H8}
\email{ballantyne, everett, murray@cita.utoronto.ca} 
\altaffiltext{1}{Current address: Department of Physics, University of
  Arizona, 1118 East 4th Street, Tucson, AZ 85721; drb@physics.arizona.edu}

\begin{abstract}
As a result of deep hard X-ray observations by \chandra\ and \xmm\ a
significant fraction of the cosmic X-ray background (CXRB) has been
resolved into individual sources. These objects are almost all active
galactic nuclei (AGN) and optical followup observations find that they
are mostly obscured Type 2 AGN, have Seyfert-like X-ray luminosities
(i.e., $L_X \sim 10^{43-44}$~ergs~s$^{-1}$), and peak in redshift at
$z \sim 0.7$. Since this redshift is similar to the peak in the cosmic
star-formation rate, this paper proposes that the obscuring material
required for AGN unification is regulated by star-formation within the
host galaxy. We test this idea by computing CXRB synthesis models with
a ratio of Type 2/Type 1 AGN that is a function of both $z$ and
2--10~\kev\ X-ray luminosity, $L_X$. The evolutionary models are
constrained by parameterizing the observed Type 1 AGN fractions from
the recent work by Barger et al. The parameterization which simultaneously best
accounts for Barger's data, the CXRB spectrum and the X-ray number
counts has a local, low-$L_X$ Type 2/Type 1 ratio of 4, and predicts a
Type 2 AGN fraction which evolves as $(1+z)^{0.3}$. Models with no
redshift evolution yielded much poorer fits to the Barger Type 1 AGN
fractions. This particular evolution predicts a Type 2/Type 1 ratio of 1--2 for
$\log L_X > 44$, and thus the deep X-ray surveys are missing about
half the obscured AGN with these luminosities. These objects are
likely to be Compton thick. The parameterization also predicts a covering
fraction approaching unity at $z \sim 1$ and $\log L_X \sim 41.5$
which would be very difficult to explain using a simple unevolving
torus model. Overall, these calculations show that the current data
strongly supports a change to the AGN unification scenario where the
obscuration is connected with star formation in the host galaxy rather
than a molecular torus alone. The evolution of the obscuration implies
a close relationship between star formation and AGN fueling, most
likely due to minor mergers or interactions.
\end{abstract}


\section{Introduction}
\label{sect:intro}
In recent years, theories of the formation and evolution of active
galactic nuclei (AGN) have been dominated by a simple geometric unification
model \citep{ant93}. It is envisaged that an optically thick obscuring
torus lies between the broad-line and narrow-line regions at a
distance of $\sim 1$~pc from the central black hole, and the
orientation to the line-of-sight of this torus is the
dominant parameter in determining the observational properties of an
AGN. The opening angle
of the torus is such that it blocks the view of the optical/UV
continuum and broad-line region for most observers, and
therefore accounts for the observed dominance of narrow-line AGN (Type
2s) over broad-line AGN (Type 1s) in the local universe
\citep{mr95}. The torus will also produce absorption in the
X-ray band, and, indeed X-ray observations generally show that optical
Type 2 AGN exhibit more soft X-ray absorption than Type 1 objects
\citep*{bas99,rms99}. However, there are exceptions to this rule
\citep[e.g.,][]{pb02,bcc03} and rapid absorption variability observed
in some local AGN \citep*{ren02,ris05} indicate that refinements to
the toroidal unification picture are required.

The geometric AGN unification model is also an important component in
explaining the shape of the cosmic X-ray background (CXRB;
\citealt{gia62}) at energies $> 1$~\kev. Between 1 and 20~\kev\ the CXRB
can be described by a power-law with a photon-index of $\Gamma \approx
1.4$ \citep{kush02,lumb02,dm04}, while most unobscured AGN are
observed to have $\Gamma=1.9$ (after taking into account the effects of
Compton reflection; \citealt{mdp93}). It was therefore postulated that the
spectrum of the CXRB can be built up by summing the spectra of many
obscured AGN over a distribution in redshift, luminosity and absorbing
column density \nh\ (\citealt{sw89}; see \citealt{gill04} for a recent
review). This picture has now been largely confirmed by very deep
X-ray observations by \chandra\ and \xmm\ that have resolved 50--90\%
of the background below 10~\kev\ into individual sources
\citep{mush00,bh05,wor05}. Interestingly, optical followup of these
objects have shown that they mostly reside at redshifts $0.7 \la z \la
1.1$ \citep[e.g.,][]{toz01,bar02,bar05} and have X-ray luminosities
typical of Seyfert galaxies ($L_X \sim 10^{43}$~erg~s$^{-1}$). This is in
contrast with the predictions of the earlier synthesis models
\citep*{mgf94,com95,gill99,plm00,gill01} which were based on the
local Type 2/Type 1 ratio, and thus expected a large number of Type 2 quasars ---
obscured high luminosity objects ($L_X > 10^{45}$~erg~s$^{-1}$) --- at
redshifts $z \sim 1-2$. Indeed, X-ray luminosity functions measured
by \citet{bar05}, \citet{ueda03}, and \citet*{has05} show that at $z
\la 1$ the space density of AGN is dominated by moderate-luminosity
Seyfert galaxies. \citet{tre04} argue, however, that an
unevolving unified model (i.e., a torus) is still consistent with the
data if the deep surveys are subject to significant selection effects
(see also \citealt{tu05}).

At the same time as the CXRB was being revealed, observations by the \textit{Hubble Space
  Telescope} showed a remarkable connection between the mass
  of the supermassive black hole in the centers of local, inactive
  galaxies and the mass of the galactic bulge
  \citep{mag98,fm00,geb00,tre02}. It now seems clear that the growth
  of both the central black hole and galaxy must be connected in a
  fundamental way, and that feedback from the accreting black hole
  and/or star forming-regions have a significant impact on the
  structure of the bulge
  \citep[e.g.,][]{sr98,fab99,kh00,wl03,mqt05,dsh05}. While the connection
  between star-formation and black hole accretion has long been
  suspected based on observations of local Seyfert 2s
  \citep[e.g.,][]{cf01,j01}, further evidence has been provided by the
  redshift distribution of the \chandra\ CXRB sources \citep{bar03,zhe04,szo04},
  which peaks at about the same redshift as the cosmic star formation
  rate \citep{mad96,ce01,lef05}. More recent CXRB synthesis models
  which allow Type 2 objects to separately evolve as infrared-bright
  star-forming galaxies \citep{fra02,gf03} can reproduce the low-$z$ peak in the redshift
  distribution, but seem to overpredict the number of obscured AGN at
  $z < 1$ \citep{gill04}.

Recent observations are providing significant pressure on the basic
assumption that an unevolving unified model can be applied for all AGN at all
redshifts. The comprehensive optical follow-up work
of \citet{bar05} on X-ray deep fields provides compelling evidence
that the fraction of Type 1 objects increases significantly with
luminosity even at redshifts as low as $z=0.1$--$0.4$. The large
numbers of obscured Seyferts at $z \la 1$ indicate galaxy formation
and evolution is still ongoing at this era, but is evidently different
than the earlier quasar epoch which seems to have a much smaller
fraction of obscured AGN \citep{per04,szo04}. Perhaps there are large
numbers of Type 2 quasars which have been missed in the deep surveys due to selection effects (i.e.,
buried quasars which release the majority of their bolometric luminosity
in the infrared; \citealt{hin95}), a
possibility that future \textit{Spitzer} surveys should
elucidate\footnote{Although very recent \textit{Spitzer} results \citep{rig05}
have shown that many of the optically faint X-ray sources, while at
higher redshift, still have Seyfert luminosities and therefore do not
contribute to evolution of the quasars.}. Alternatively, the lack of
Type 2 quasars
may be explained by invoking the 'receding torus' model \citep{law91,simp05}, where,
because of the effects of dust sublimation, the inner radius of the
torus is increased at higher luminosity. Another possibility is that
radiation pressure flattens the torus in high luminosity objects
\citep{kk94}. Both these explanations, since they assume no connection
between the AGN and the host galaxy, should be independent of redshift, but
Figure~19 in the paper by \citet{bar05} also shows evidence that the
broad-line AGN fraction decreases with redshift at a given
luminosity. These authors find that at the redshift where the bulk of
the CXRB is produced ($z \sim 0.7-1$),
the Type 1 fraction at a luminosity of $10^{43}$~erg~s$^{-1}$ is only
$\sim$20\% as compared to 70--80\% at $z=0.1-0.4$. In addition, the
low redshift ($z < 0.15$) optical Type 1 and 2 luminosity functions
constructed by \citet{hao05} from nearly 3000 AGN in the Sloan Digital Sky
Survey find a Type 2/Type 1 ratio of approximately unity at low
Seyfert luminosities, with Type 1s dominating at higher luminosities
(see also \citealt{heck05}). These results all imply that the
obscuring material responsible for blocking the AGN continuum in
Seyfert 2s evolves with redshift, as well as luminosity. Moreover, if the
Sloan results are accurate the 
absorbing medium must evolve so that it increases its covering
fraction by a factor of 3--4 from $z \sim 0$ to $z \sim 1$. This
evolution exactly tracks that of the star formation rate in the universe.

The central idea proposed here is that the obscuration important for
providing the correct shape of the CXRB is regulated by and controlled via
active star-formation within the galaxy. The actual processes
responsible for the obscuration are unknown, but it may be associated
with star-formation on many-hundred pc scales, or be brought in closer
to the nucleus by a starburst disk \citep*{tqm05}. Thus, the Type
2/Type 1 ratio does evolve with redshift and peaks at $z \sim 1$,
where both the star-formation rate is a maximum and the majority of
the obscured AGN discovered by the deep surveys are found. Similar
ideas have been discussed before \citep{fab98,fra99}, but here we use
the latest observational constraints on the CXRB spectrum, source
number counts and Type 2/Type 1 ratio ($\equiv R$) to determine what
kind of AGN evolution is possible and if it is consistent with the
above hypothesis.

We begin by describing in the next section the calculation of the CXRB
spectrum and number counts, including assumptions on the AGN spectral
model and evolution of $R$. Section~\ref{sect:results} presents the
results of the computations and compares them against the data. We
discuss the implications of the results for the unified model and
models of galaxy formation/evolution in Sect.~\ref{sect:discuss}, and
then present our conclusions in Sect.~\ref{sect:concl}. 

This paper assumes the standard \textit{WMAP} and \textit{Boomerang}
$\Lambda$-dominated cosmology: $H_0=70$~km~s$^{-1}$~Mpc$^{-1}$, $\Omega_{\Lambda}=0.7$, and $\Omega_{m}=0.3$ \citep{spe03,mct05}.

\section{CXRB Synthesis Model}
\label{sect:cxrb}
\subsection{Formalism}
\label{sub:form}
The calculation of the CXRB spectrum and number counts follows the
standard method outlined in earlier works \citep[e.g.,][]{com95,plm00}. 

The spectral intensity of the CXRB (in keV cm$^{-2}$ s$^{-1}$
keV$^{-1}$ str$^{-1}$) at energy $E$ is given by 
\begin{equation}
I(E) = {c \over H_0} \int_{z_{\mathrm{min}}}^{z_{\mathrm{max}}} \int_{\log L_{X}^{\mathrm{min}}}^{\log
  L_{X}^{\mathrm{max}}} {d\Phi(L_X,z) \over d\log L_X} {S_{E}(L_X,z) d_l^2 \over (1+z)^2
  (\Omega_m (1+z)^3 + \Omega_{\Lambda})^{1/2}} d\log L_X dz,
\label{eq:cxrb}
\end{equation} 
where $d\Phi(L_X,z)/d\log L_X$ is the X-ray luminosity function for AGN
(in Mpc$^{-3}$), $S_{E}(L_X,z)$ is the absorbed observed-frame spectrum of an AGN (in keV
cm$^{-2}$ s$^{-1}$ keV$^{-1}$) with intrinsic luminosity $L_X$ at
redshift $z$, and $d_l$ is the luminosity distance to redshift $z$:
\begin{equation}
d_l = (1+z) {c \over H_0} \int_{0}^{z} {dz' \over (\Omega_m (1+z')^3 +
  \Omega_{\Lambda})^{1/2}}.
\label{eq:dl}
\end{equation}

A similar expression is used to calculate the number of sources per
square degree with flux (defined in some energy band) greater than $F$, $N(>F)$
\begin{equation}
N(>F)= {K_{\mathrm{str}}^{\mathrm{deg}}c \over H_0}
  \int_{z_{\mathrm{min}}}^{z_{\mathrm{max}}} \int_{\mathrm{max}(\log
  L_{X}^{\mathrm{min}},\, \log L^F_X)}^{\log L_{X}^{\mathrm{max}}}
  {d\Phi(L_X,z) \over d\log L_X} {d_l^2 \over (1+z)^2 (\Omega_m (1+z)^3 +
  \Omega_{\Lambda})^{1/2}} d\log L_X dz, 
\label{eq:counts}
\end{equation} 
where $K_{\mathrm{str}}^{\mathrm{deg}}=3\times 10^{-4}$ is the
conversion factor from str$^{-1}$ to degrees$^{-2}$, $\log L^F_X$
is the rest-frame luminosity corresponding to observed-frame flux $F$
at redshift $z$. 

\subsection{Ingredients}
\label{sub:ingredients}
To generate a CXRB model with the above equations, a number of key
ingredients must be specified.
\subsubsection{The Luminosity Function}
\label{subsub:xlf}
For the normalization and evolution of the space density of AGN, we
use the hard X-ray luminosity function (HXLF) of \citet{ueda03}. It is
defined for the absorption corrected luminosity in the rest-frame
2--10~\kev\ band. As it is absorption-corrected, it describes the
density and evolution of all detected AGN, both Type 1 and 2.

\citet{ueda03} found that the evolution of the luminosity function was
best described using a luminosity-dependent density evolution (LDDE)
model such that the lower luminosity objects peaked in density at
lower redshift than sources with higher luminosities. In contrast,
\citet{bar05} found that pure luminosity evolution was the best fit to
their hard X-ray luminosity function at $z < 1$. However, recent work by \citet{has05} 
and \citet{laf05} have both found that LDDE best describes AGN evolution
from $z \sim 0$ to $z \sim 4-5$. In this paper, we assume the LDDE
model and parameters of \citet{ueda03}.

Following \citet{ueda03} and \citet{tu05}, we take $z_{\mathrm{min}}=0$,
$z_{\mathrm{max}}=5$, $\log L_{X}^{\mathrm{min}}=41.5$, and $\log
L_{X}^{\mathrm{max}}=48$, where from this point on the luminosity $L_X$ refers to
2--10~\kev\ luminosity. 

\subsubsection{Spectrum}
\label{subsub:spectrum}
Consistent with using the \citet{ueda03} HXLF where there is no
distinction between the Type 1 or Type 2 objects, a single spectral
model is assumed for all the AGN in the calculation. \citet{she05}
have recently shown that there is little change in the X-ray spectral
shape of AGNs over a large range of luminosity and redshift. The
spectrum defined here
consists of a power-law with photon-index $\Gamma=1.9$ and high-energy
rollover at 375~\kev, augmented by Compton reflection. The
\textsc{pexrav} model of \citet{mz95} is used to calculate the
reflection spectrum with a reflection fraction of unity, solar
abundances, and an inclination angle of 63~degrees. This model is very
similar to ones assumed in other CXRB synthesis calculations
\citep[e.g.,][]{tu05}. The spectrum was calculated from 0.1 to
900~\kev\ in 1000 logarithmically spaced steps.

\subsubsection{K-correction}
\label{subsub:Kcorrect}
The following expression is used to convert from 2--10~\kev\ flux,
$F_{\mathrm{2-10}\ \mathrm{keV}}$, to luminosity \citep[e.g.,][]{hogg02}:
\begin{equation}
L_X^{\mathrm{2-10}}={4\pi d_l^2 F_{\mathrm{2-10}\ \mathrm{keV}} \over (1+z)}.
\label{eq:Kcorrect}
\end{equation}

\subsubsection{\nh\ Distribution}
\label{subsub:nh}
As has been standard over the last few years, Type 2 objects are
defined here to be those which are seen through columns $\log
N_{\mathrm{H}} \geq 22$, with Type 1s being less absorbed. Once the
ratio of absorbed to unabsorbed objects ($R$) is calculated for a
given $z$ and $L_X$ (see \S~\ref{sub:ratio}), the \nh\ distribution for
the obscured and unobscured AGN must be defined.  However, the distribution
of absorbing columns for AGN is unknown except for locally, where
\citet{rms99} measured it for nearby luminous Seyfert~2s.

Ten \nh\ bins are defined in the calculation: $\log N_{\mathrm{H}}=20,
20.5, \ldots, 24, 24.5$. The implicit assumption is that objects which
are severely Compton-thick with $\log N_{\mathrm{H}} \geq 25$ will not
significantly contribute to the observed CXRB. In lieu of any other
information, we first make the simplest choice and assume a flat \nh\
distribution for both Type 1 and Type 2 sources. Computations were
also performed with the \citet{rms99} distribution
where 75\% of all Type 2s have $\log N_{\mathrm{H}} \geq 23$ and half
are Compton thick.

\begin{figure}
\epsscale{0.85}
\plotone{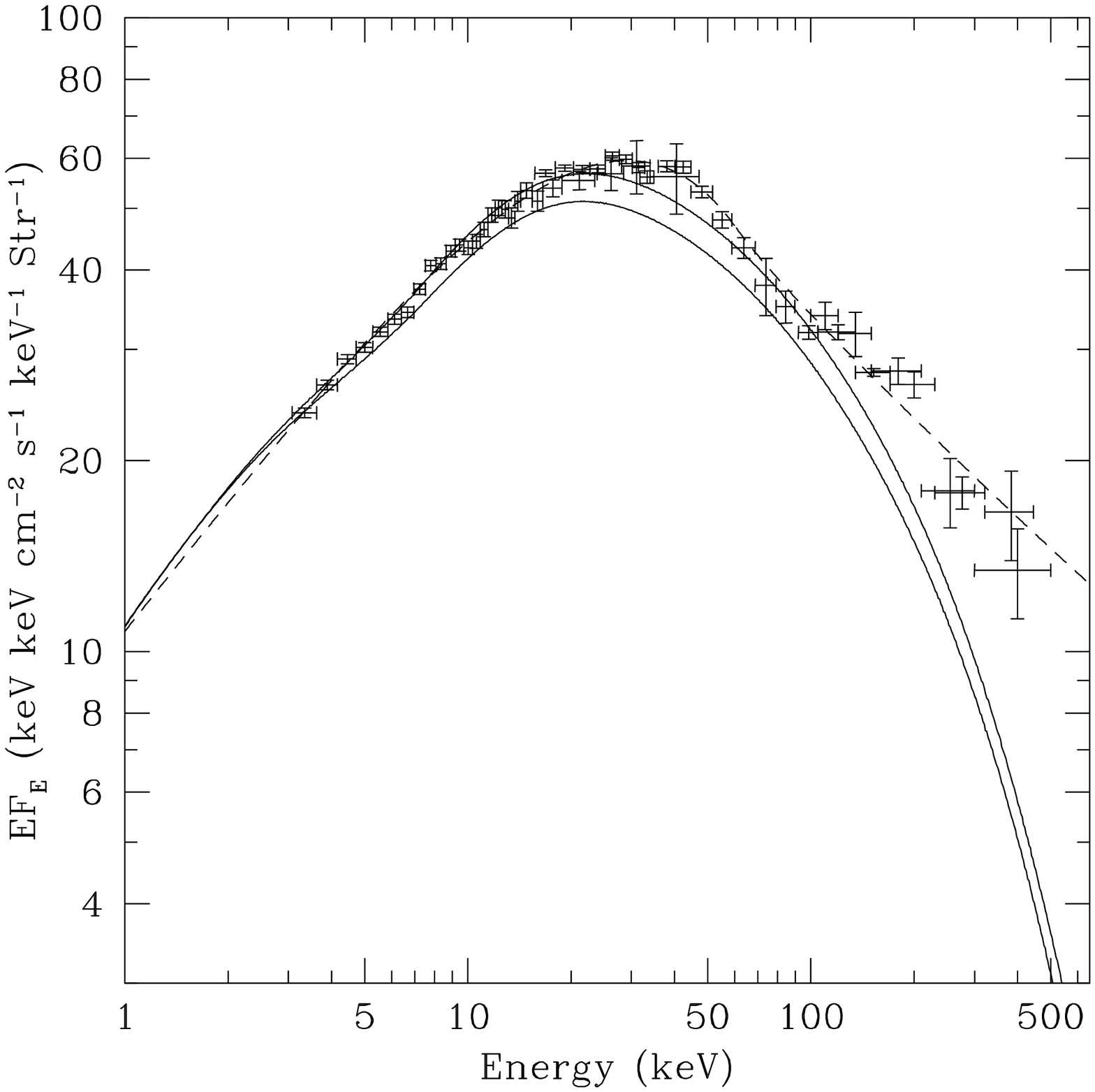}
\caption{The CXRB spectrum between 1 and 600~\kev. The data points
  are from the HEAO-1 experiment \citep{gru99}, shifted upwards by 40\% to
  more closely match newer measurements \citep{dm04}. The dashed line is the
  analytical fit to the HEAO-1 data provided by \citet{gru99}. It has also
  been shifted upwards by 40\%. The upper solid curve is a model of
  the CXRB produced using the same parameters as \citet{tu05}. The major
  differences from our model are a reflection fraction of 2, a
  high-energy cutoff of 300~\kev\ and a constant $R=3$ for all $L_X$ and
  $z$. The lower solid curve is identical to the previous one except
  that the effects of Compton-thick absorbers are properly
  included. This has an important effect on the shape of the spectrum,
  particularly around the peak of the intensity at 20--40~\kev. Both
  curves have had their normalizations increased by a factor of 1.5 to
  match the data at low energy. }
\label{fig:tu}
\end{figure}

For $\log N_{\mathrm{H}} \leq 23$, the solar
abundance photoelectric cross-sections of \citet{mcm83} are used to modify the
spectrum. At higher columns, where Compton scattering becomes
important within the absorber, we extracted the results plotted by
\citet{matt99} and divided them by their incident spectrum ($\Gamma=2$
power-law with cutoff energy of 500~\kev) to obtain the
absorption/transmission curve in our energy range. We then applied the
appropriate curve to our model spectrum to account for Compton-thick
absorbing columns. The effects of correctly including the
absorption and transmission properties of Compton-thick columns are
illustrated in Figure~\ref{fig:tu} where we reproduce the results of
\citet{tu05}. Including the proper treatment of Compton
thick absorbers produced a poorer fit for this set of parameters.

\subsection{The Evolving Type 2/Type 1 Ratio}
\label{sub:ratio}
The central hypothesis in this work is that the Type 2/Type 1 ratio, $R$,
will increase with redshift until $z \sim 1$. In addition, $R$ will decrease with
$L_X$ as seen in many recent observational campaigns. Since the leading
explanations for this effect are based on radiative processes
associated with the
central engine, we assume that there is no redshift dependence for
this $L_X$-evolution. There is still, however, immense freedom in how the evolution
of $R(L_X,z)$ can be described, and as can be seen from previous work
\citep[e.g.,][]{gill99,gf03,tu05}, the CXRB is not very constraining on its own. 

We therefore make use of the results presented by \citet{bar05} in
their Figure~19 which shows the observed fraction of Type~1 objects 
as a function of 2--8~\kev\ luminosity in three different redshift
bins: $0.1-0.4$, $0.4-0.8$, and $0.8-1.2$. While the exact parameterization of $R(L_X,z)$ 
remains unconstrained, Barger's data provides landmarks to which the
model can be pinned. Ideally, we would be able
to compute the Type~1 fraction from a self-consistent model of star
formation and black hole growth from redshift 1 to zero. However, such a
complicated calculation is not necessary for this study. At this
stage, it suffices to determine if redshift evolution of $R$ is needed
to explain the observations, and, if so, to quantify the degree of the evolution. We are thus searching for the best
description of $R(L_X,z)$ that simultaneously accounts for Barger's
data, the CXRB spectrum, and the X-ray number count distributions (in
multiple bands).

In practice, we do not compute $R(L_X,z)$, but the related quantities
$f_2(L_X,z)$ and $f_1(L_X,z)$, the fraction of Type 2s and Type 1s,
respectively. These three parameters are related by
\begin{equation}
f_2(L_X,z)={R(L_X,z) \over 1+R(L_X,z)}
\label{eq:f2}
\end{equation}
and
\begin{equation}
f_1(L_X,z)=1-f_2(L_X,z).
\label{eq:f1}
\end{equation}
Four different parameterizations were used to explore the evolution of
the AGN Type 2/Type 1 ratio from $z=0-1$. In addition to the assumptions outlined
above, these particular forms were chosen solely as
good candidates for accounting for the observed data from
\citet{bar05}, i.e., to smoothly run from 0--1 over the observed range
of $L_X$. The parameterizations are:
\begin{equation}
f_1=K \exp \left (L_X \over {10^{\beta} (1+z)^{\alpha}} \right ),
\label{eq:expmodel}
\end{equation}
\begin{equation}
f_1=K(1+z)^{-\alpha} \left (L_X \over 10^{41.5} \right )^{\beta},
\label{eq:plawmodel}
\end{equation}
\begin{equation}
f_2=K(1+z)^{\alpha} (\log L_X)^{-\beta},
\label{eq:loglmodel}
\end{equation}
and
\begin{equation}
f_2=K(1+z)^{\alpha} \cos^2 \left( {\log L_X -41.5 \over \beta} \right ).
\label{eq:cosmodel}
\end{equation}
Aside from $\alpha$ and $\beta$ the only other free parameter is the
normalization $K$. This is set by defining $R$ at $z=0$ and $\log
L_X=41.5$, denoted by $R_0$. At redshifts $\geq 1$ the fractions $f_2$
or $f_1$ are fixed at their $z=1$ values, mimicking the roughly flat
star-formation rate during this era \citep{ce01}. If the value of
$f_1$ or $f_2$ ever became negative or $> 1$ then it was forced
to zero or one, respectively.

For every $R_0=1,2,\ldots,9,10$, we search for the best $\alpha$ and
$\beta$ in each of the four parameterizations by $\chi^2$ fitting to
the Barger data. Equations~\ref{eq:expmodel}-\ref{eq:cosmodel} are
evaluated at the midpoints of the three redshift bins used by
\citet{bar05}, namely, $z=0.25, 0.6$ and $1.0$. The values of $L_X$ are
taken directly from the Barger data, and we ignore the small
difference between 2--8~\kev\ and 2--10~\kev\ luminosities (17\% when
$\Gamma=1.9$). The errorbars on each datapoint are obtained from their
plot. If only one errorbar is visible then we assume they are
symmetric. The $\log L_X \approx 42.8$ data point in the $z=0.8-1.2$ bin
and the $\log L_X \approx 42.3$ point in the $z=0.4-0.8$ bin are omitted
in the $\chi^2$ calculation because they were an upper-limit or had
very small errorbars. The $\chi^2$ calculation took into account the
data from all three redshift bins, thus we are searching for the model
that can best account for the observed redshift and luminosity
evolution over the entire range in $z$ and $L_X$. There are 13 data
points and 2 free parameters, resulting in 11 degrees of freedom (d.o.f.).

The minimum $\chi^2$ is searched for by iterating over a large
$\alpha$ and $\beta$ parameter space.  Once the best fit $\alpha$ and
$\beta$ are known for each $R_0$, the evolution
equations~\ref{eq:expmodel}-\ref{eq:cosmodel} are inserted into the
CXRB machinery described in \S~\ref{sub:form}. Predictions of the CXRB
spectrum and number counts are then computed for the different
evolutions and compared with the observations. As in previous work
\citep{gf03}, the normalization of the predicted spectrum is
multiplied by a constant in order to fit the observed data
(whose normalization is also uncertain; \citealt{dm04}).

\section{Results}
\label{sect:results}

\subsection{$z$ and $L_X$ evolution}
\label{sub:basic}
The best fit values of $\alpha$ and $\beta$ for all four evolution
models are shown in Table~\ref{table:constantalpha}, along with the
computed $\chi^2$ and the corresponding null-hypothesis probability
$P_{\chi^2}$. The power-law parametrization (Eq.~\ref{eq:plawmodel})
seems to provide the best fit to the Barger data for every assumed
$R_0$, with some very low values of $\chi^2$ obtained when $R_0 >
4$. The cosine model (Eq.~\ref{eq:cosmodel}) is the next best
parameterization, with good fits obtained for every $R_0 > 1$. The
exponential model (Eq.~\ref{eq:expmodel}) only finds very good fits
at large values of $R_0$, while the $\log L_X$ power-law model
(Eq.~\ref{eq:loglmodel}) results in $\chi^2/$d.o.f.$\gtrsim 1$. 

In order to use these values of $\alpha$ and $\beta$ to predict the
CXRB properties, an \nh\ distribution must be assumed. In
\S~\ref{subsub:simple} we use a simple prescription where the columns
of Type II AGN are spread uniformly between $\log N_{\mathrm{H}}=22$
and $24.5$, and similarly for Type I AGN between $\log
N_{\mathrm{H}}=20$ and $21.5$. Results assuming the \citet{rms99} \nh\
distribution are presented in \S~\ref{subsub:rms}. 

\subsubsection{Simple \nh\ Distribution}
\label{subsub:simple}
\begin{figure}
\epsscale{0.95}
\plottwo{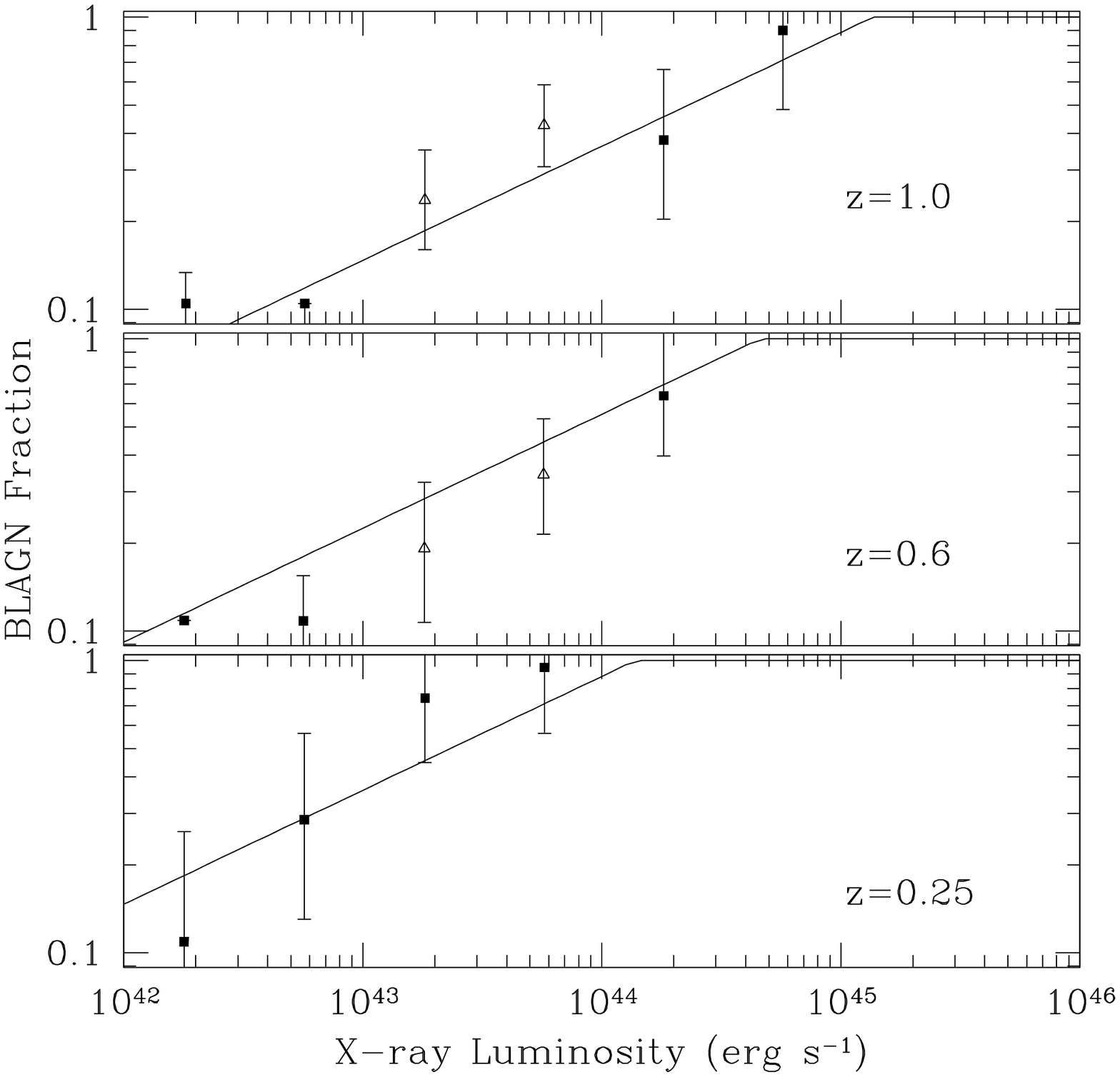}{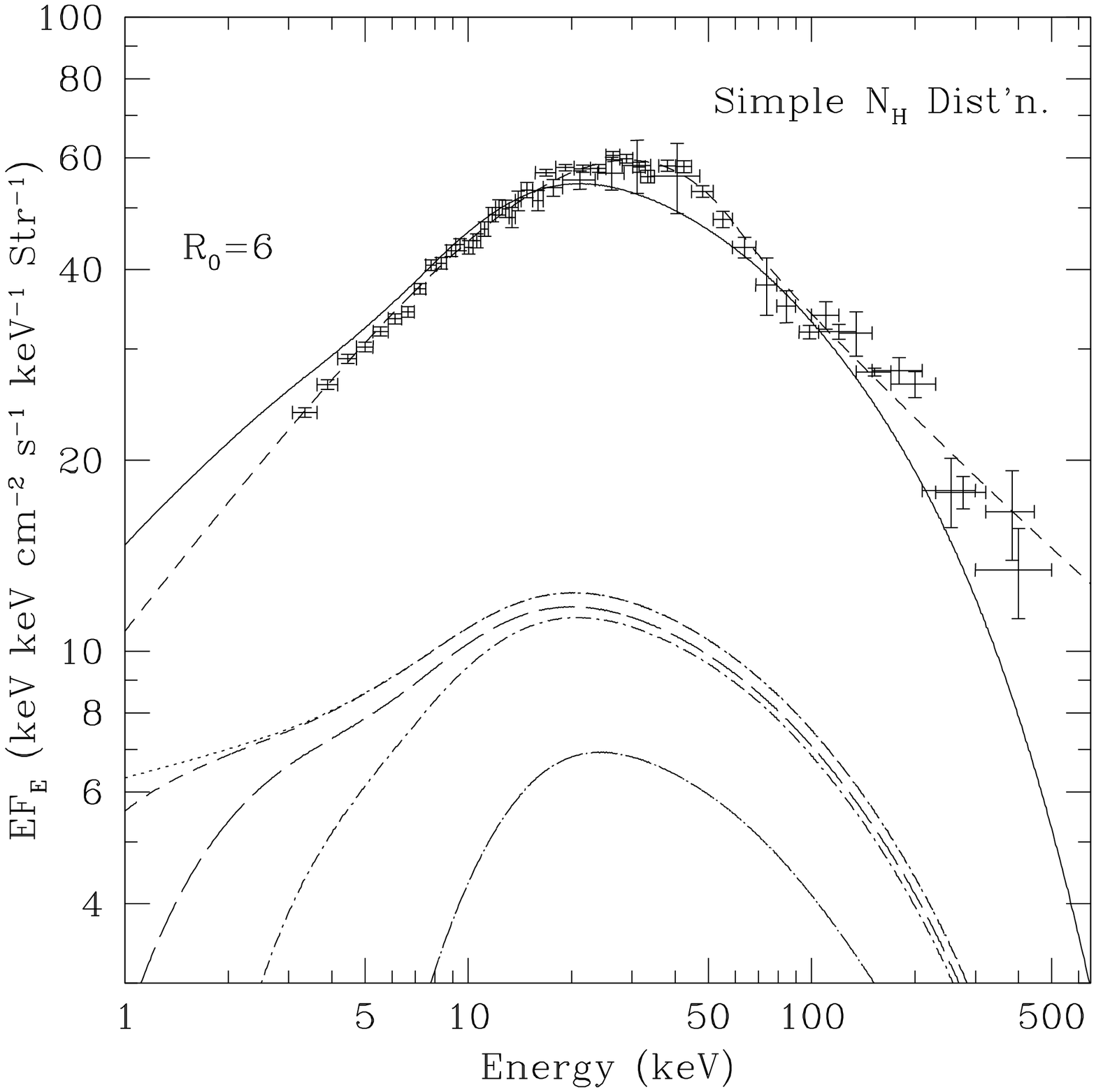}
\epsscale{0.475}
\plotone{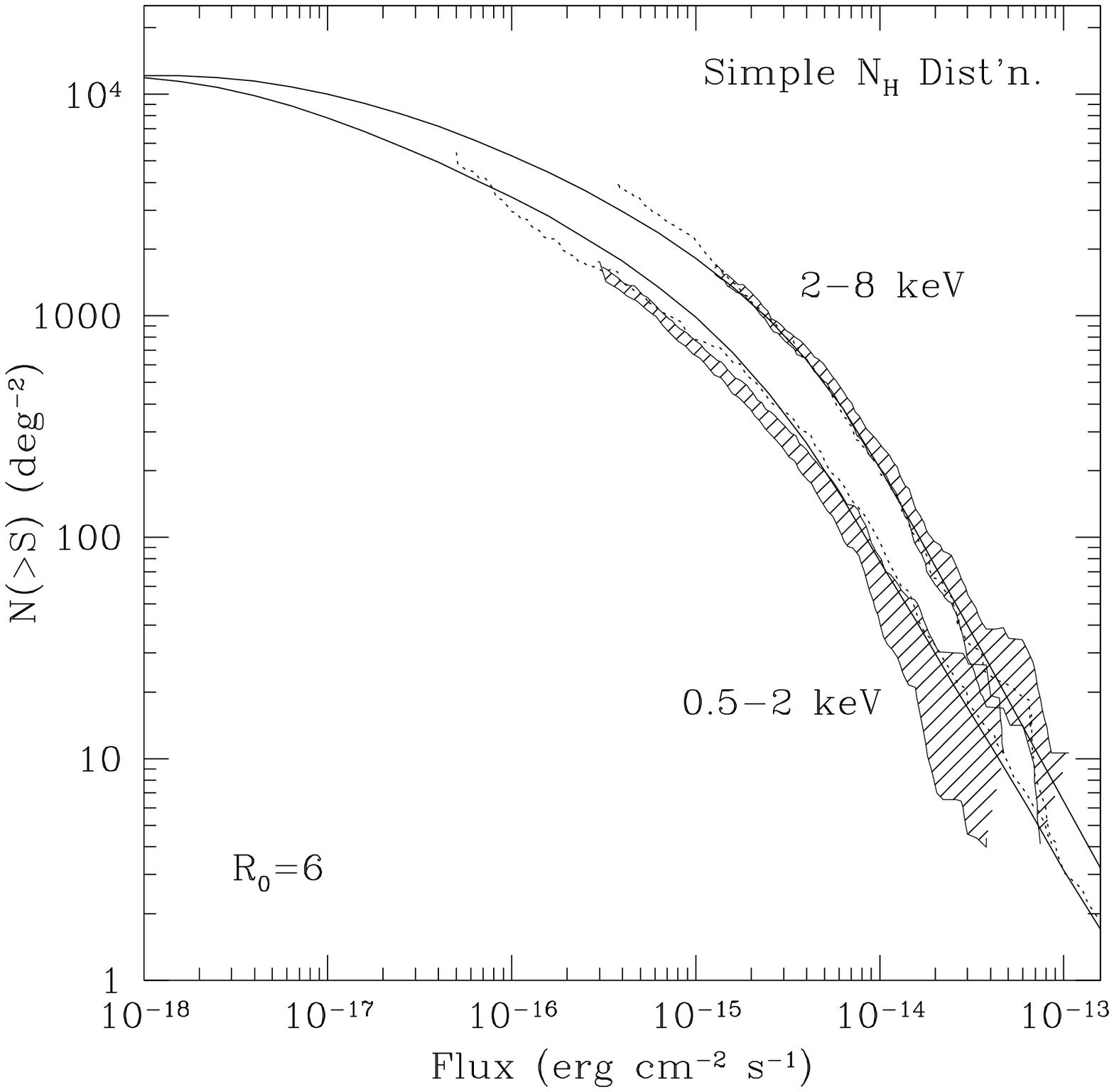}
\caption{Results from the power-law parameterization (Eq.~\ref{eq:plawmodel})
  when $R_0=6$. \emph{(Upper-left panel)} The solid line plots the
  predicted fraction of Type 1 or broad-line AGNs (BLAGNs) at $z=0.25, 0.6$ and
  $1.0$. The points are the observed values from \citet{bar05}. Open
  triangles represent sources which contribute the majority of the
  CXRB \citep{ueda03,bar05}. This particular parameterization is an extremely
  good fit to the Barger data with
  $P_{\chi^2}=0.8$. \emph{(Upper-right panel)} The CXRB spectrum
  predicted by this evolutionary model. The lines and points are as in
  Fig.~\ref{fig:tu}. In addition, the contribution from sources with
  $\log N_{\mathrm{H}}=20-21$ (dotted line), $21-22$ (short dashed
  line), $22-23$ (long dashed line), $23-24$ (dot-short dashed line),
  and $24-24.5$ (dot-long dashed line) are also plotted. \emph{(Lower
  Panel)} The solid lines show the predictions for the 0.5--2~\kev\
  and 2--8~\kev\ X-ray number counts using this parameterization for the AGN
  Type 2/Type 1 ratio. The hatched regions are the observed
  distributions from the Extended \chandra\ Deep Field South
  \citep{leh05}. The dotted line in the 2--8~\kev\ band is from the
  \chandra\ Deep Field South \citep{ros02}, while the dotted line in
  the 0.5--2~\kev\ band is from \citet{has05}.}
\label{fig:6to1}
\end{figure}

While the normalization of the CXRB is still uncertain \citep{dm04}, the
spectral slope is well constrained to be $\Gamma \approx 1.4$ above
1~\kev\ \citep[e.g.][]{kush02}. To provide a quantitative estimate of how well our evolutionary
models fit the CXRB spectrum, we computed $\Gamma$ between 1 and
10~\kev\ and the results are listed in the penultimate column of
Table~\ref{table:constantalpha}. Interestingly, the power-law
evolutionary model, which gives the best fit to the Barger data, fails
badly in recovering the shape of the CXRB spectrum. This is
illustrated in Fig.~\ref{fig:6to1} for the case of $R_0=6$. We find
that there are far too few absorbed sources in this model to
successfully account for the shape of the observed CXRB. In fact, we
find that, in general, the closer one fits the \citet{bar05} data, the poorer the
fit to the CXRB spectrum.

\begin{figure}
\epsscale{1.0}
\plottwo{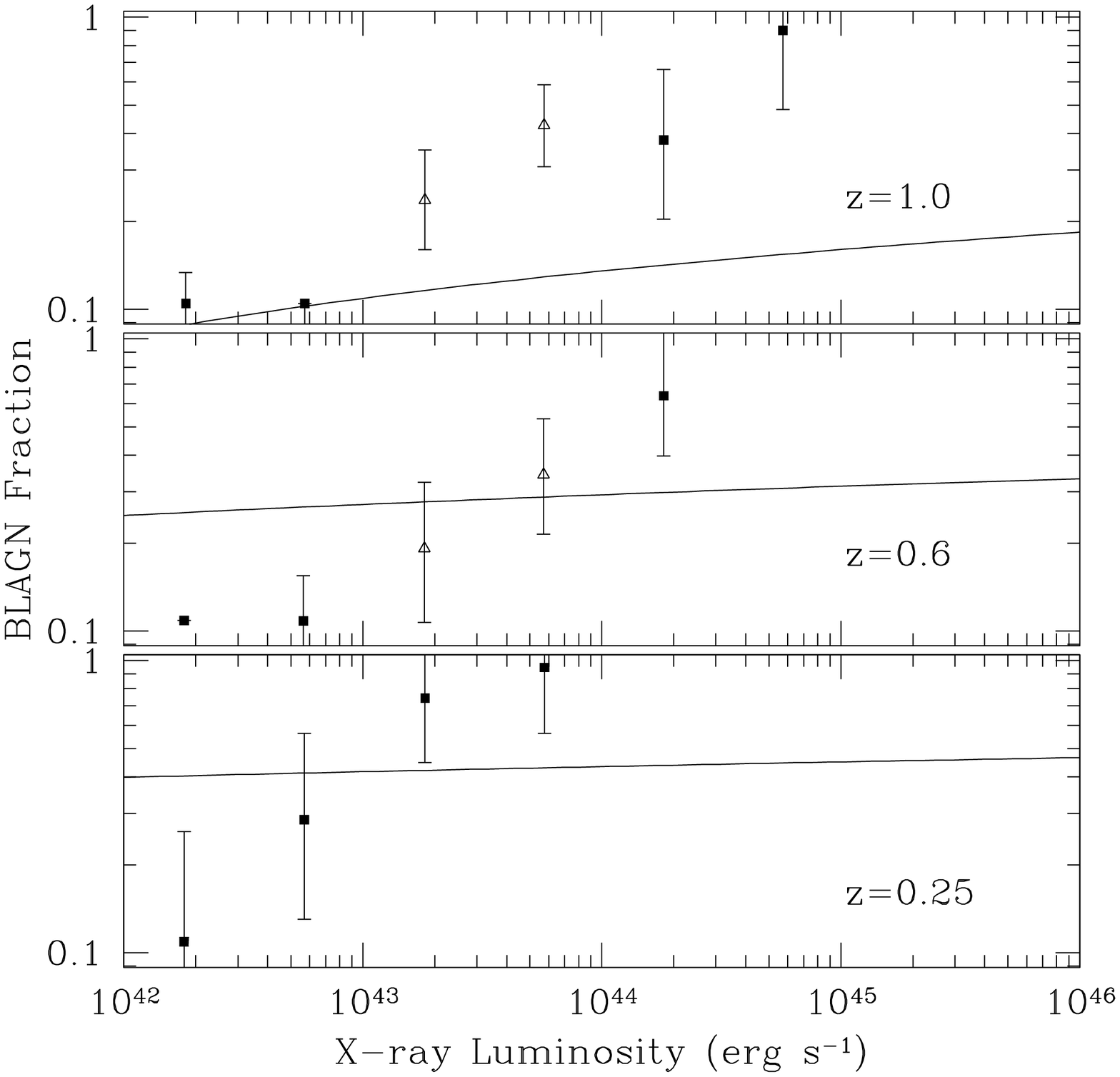}{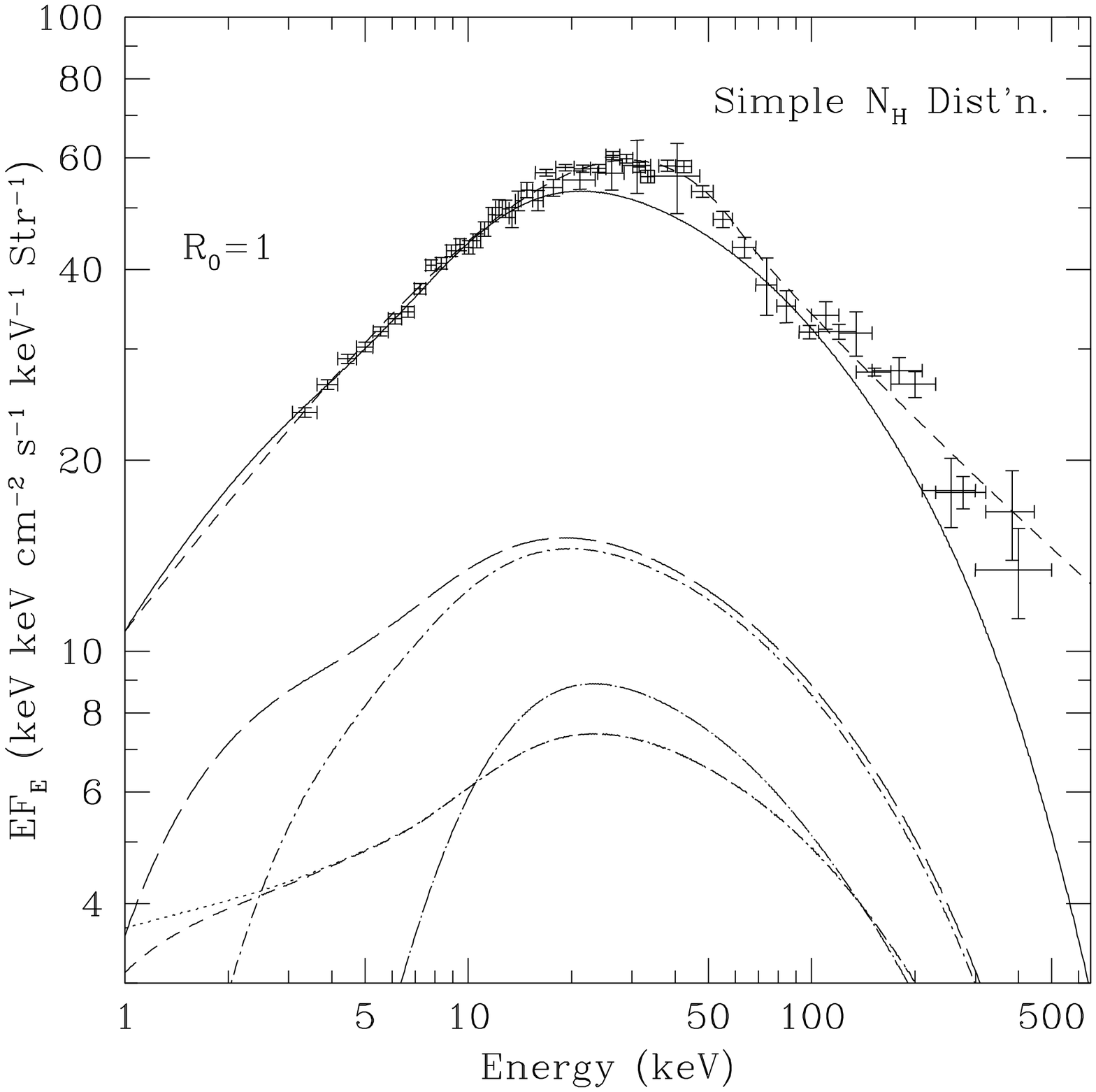}
\epsscale{0.5}
\plotone{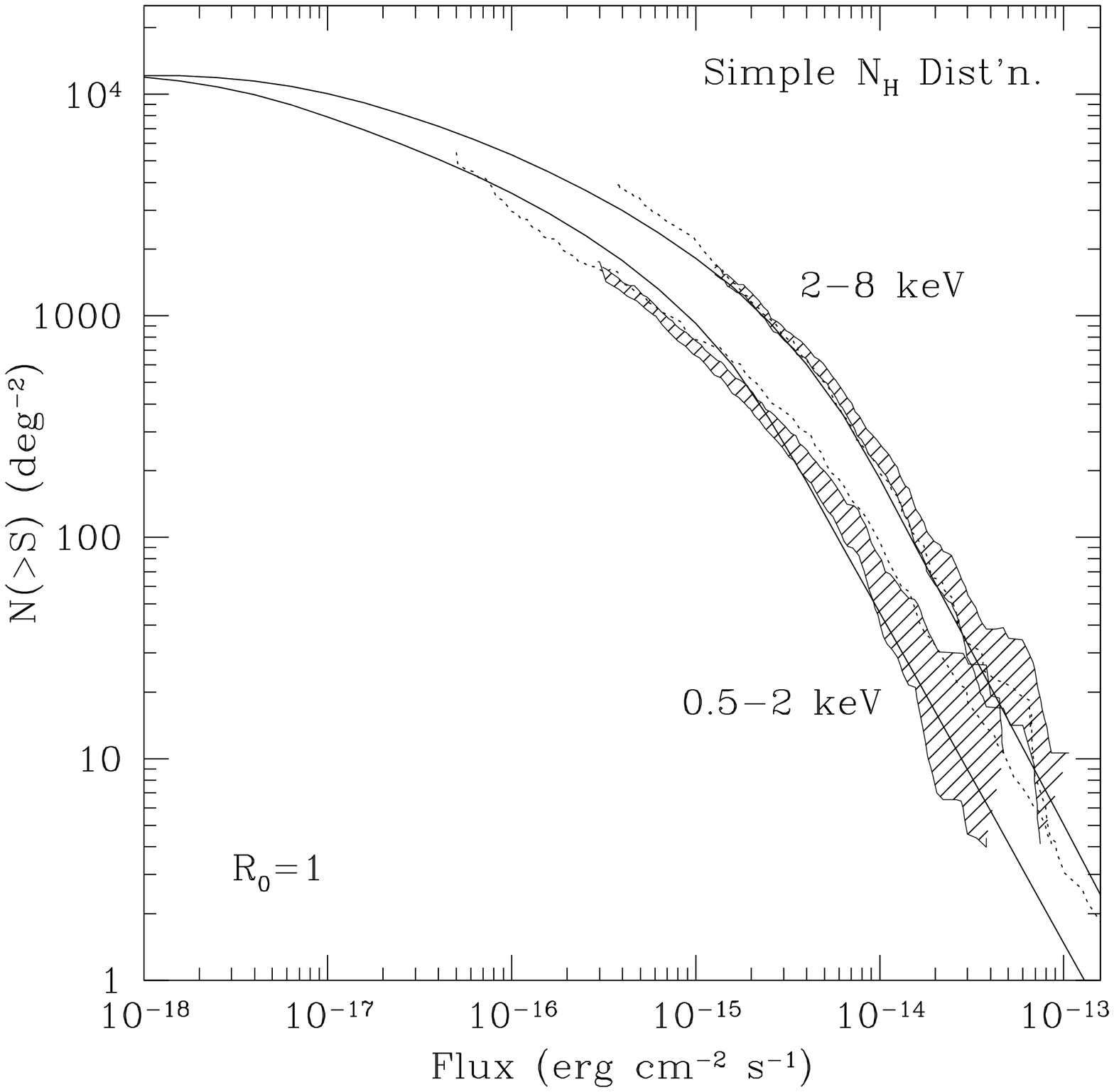}
\caption{Results from the $\log L_X$ power-law parameterization (Eq.~\ref{eq:loglmodel})
  when $R_0=1$. Plots are as in Fig.~\ref{fig:6to1}. 
  The fit to the Barger data is poor with $P_{\chi^2}=2.8\times
  10^{-3}$.}
\label{fig:1to1}
\end{figure}

The best match to the observed CXRB was found with the $\log L_X$
power-law model (Eq.~\ref{eq:loglmodel}), the one that provided
decent, but not spectacular, fits to Barger's data ($P_{\chi^2} \sim 0.2-0.4$). Figure~\ref{fig:1to1} plots the results from
this evolutionary model assuming $R_0=1$, the value pointed to by the recent
Sloan analysis \citep{hao05,heck05}. This model can successfully
reproduce the properties of both the CXRB spectrum and the number
counts, but produces a very poor fit to Barger's Type 1
fractions. Indeed, the model, which requires very strong $z$ evolution to fit
the CXRB, predicts that the X-ray surveys are missing many low-$L_X$
Type~1 objects at $z \leq 0.6$ as well as many high-$L_X$ absorbed objects
at all $z$, both by factors of 2--5. As a consequence, Type 2 quasars
are predicted to be 4--5 times as numerous as Type 1 quasars at $z
\gtrsim 1$, a result that does not seem to be borne out by recent
observations \citep{mart05}.

\begin{figure}
\epsscale{1.0}
\plottwo{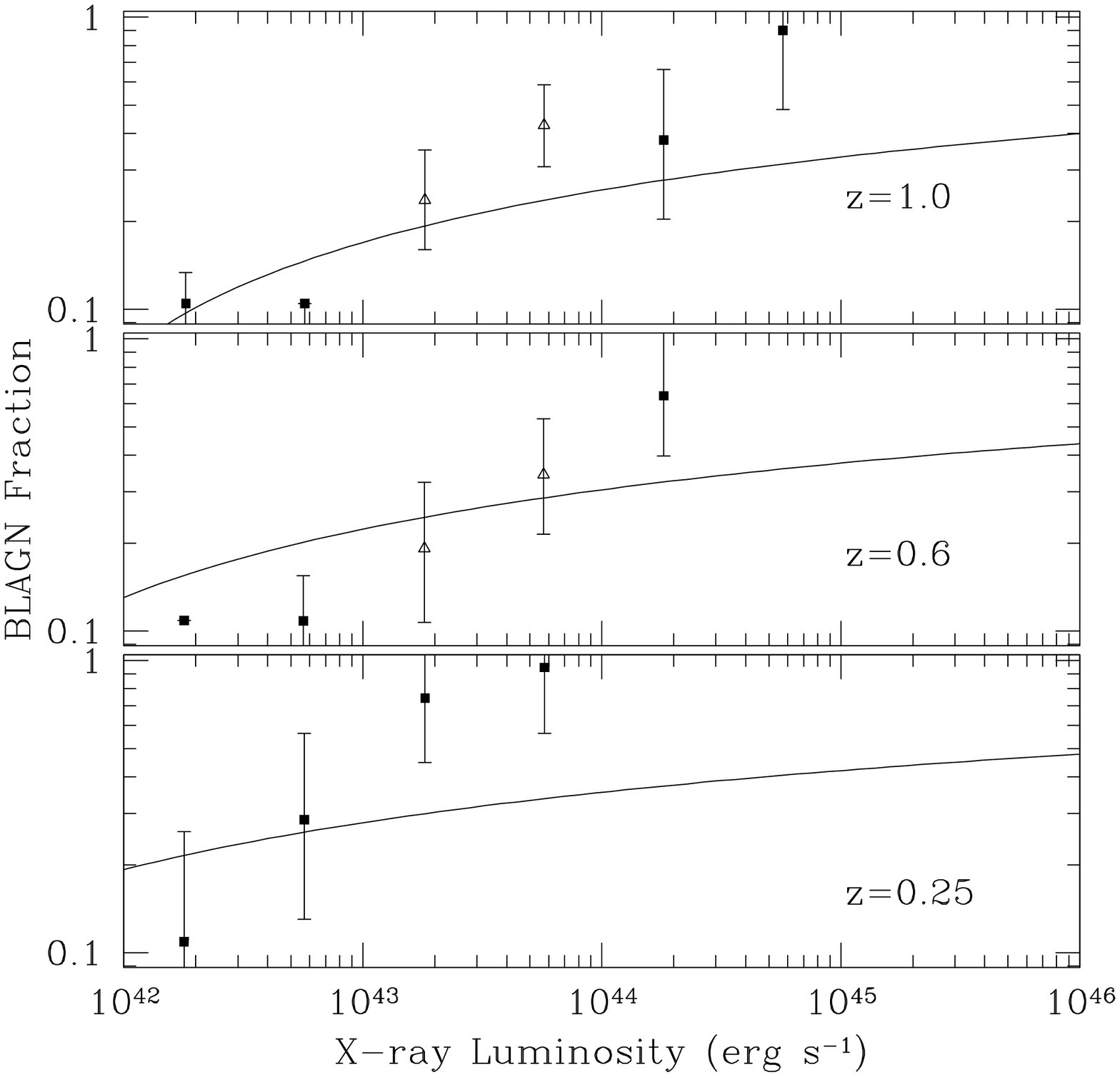}{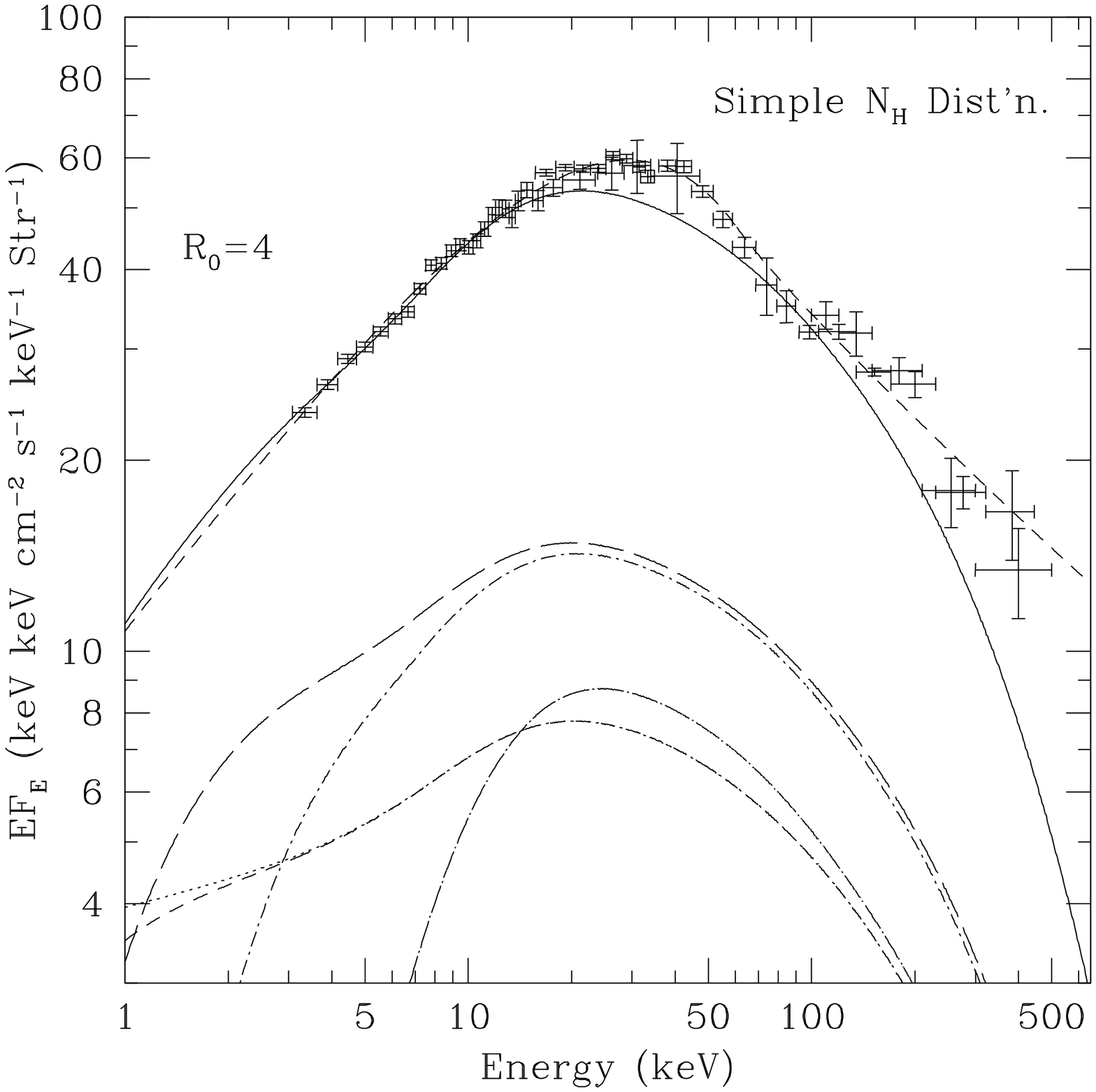}
\epsscale{0.5}
\plotone{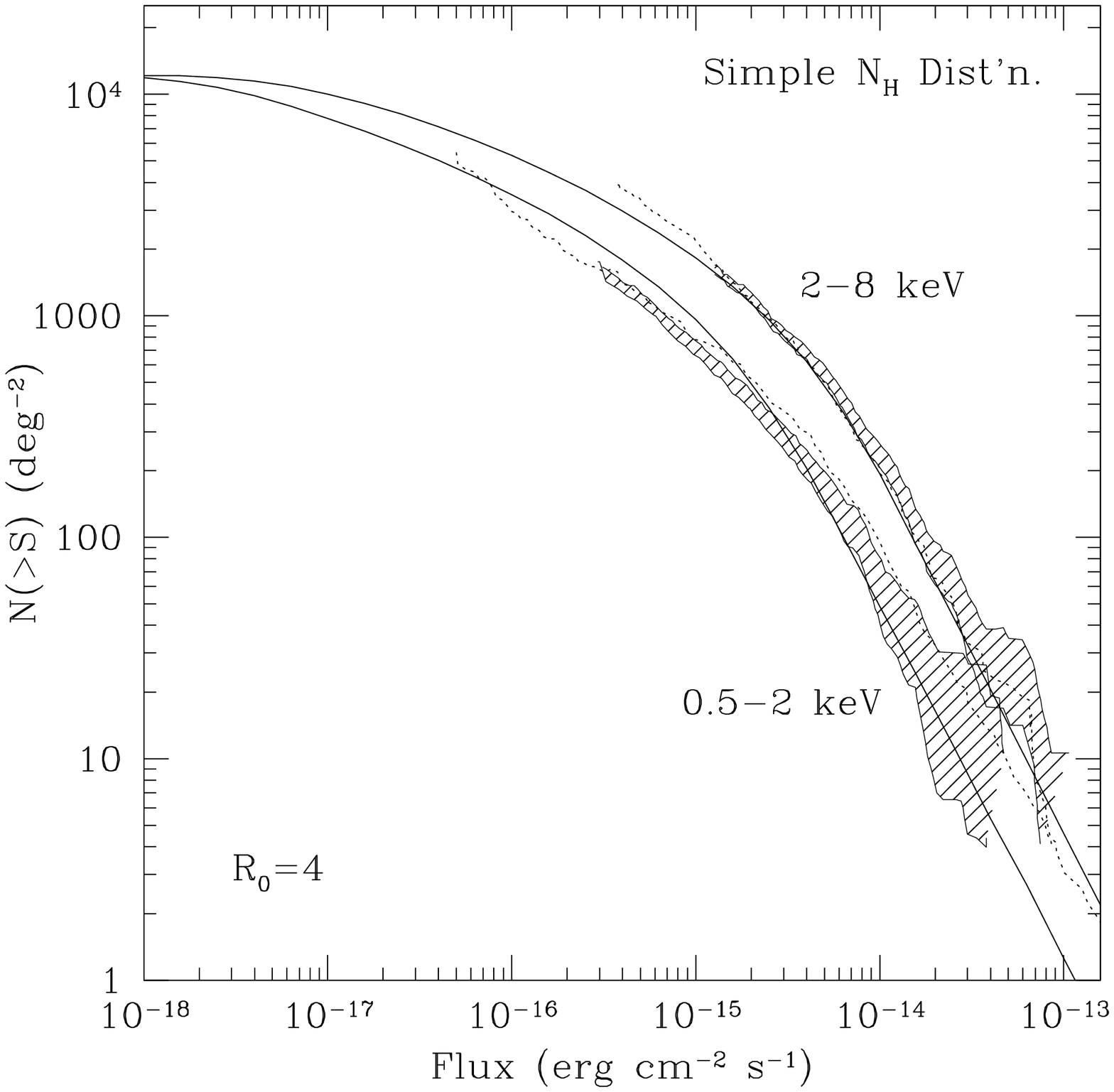}
\caption{Results from the $\log L_X$ power-law parameterization (Eq.~\ref{eq:loglmodel})
  when $R_0=4$. Plots are as in Fig.~\ref{fig:6to1}. 
  The fit to the Barger data has a $P_{\chi^2}=0.24$.}
\label{fig:4to1}
\end{figure}

\begin{figure}
\epsscale{0.85}
\plotone{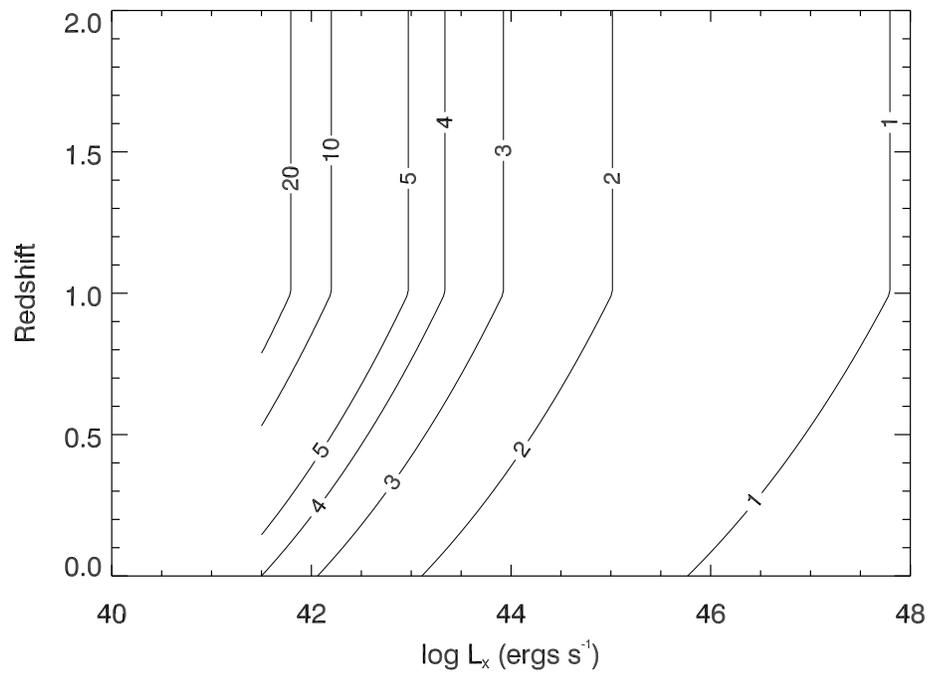}
\caption{A contour plot of $R$, the AGN Type 2/Type 1 ratio, as a
  function of $\log L_X$ and $z$ predicted from the $R_0=4$ model of
  Eq.~\ref{eq:loglmodel} (Fig.~\ref{fig:4to1}). Recall that the
  evolution in redshift was halted at $z=1$.}
\label{fig:contour}
\end{figure}

Figure~\ref{fig:4to1} represents a compromise solution. Here,
we plot the results of the $\log L_X$ power-law model with $R_0=4$,
similar to the classic ratio of
\citet{mr95}. Table~\ref{table:constantalpha} tells us that the fit to
Barger's data is adequate, and that the 1--10~\kev\ slope is $\Gamma=1.418$,
close to the observed value. The predicted number counts seem
consistent with the observations from the \chandra\ deep fields. In
this case, the prediction is that the X-ray surveys are missing $\sim
50$\% of the absorbed high-$L_X$ objects. In
Figure~\ref{fig:contour} we plot contours of $R(L_X,z)$ predicted by
this evolution model between $z=0-2$. The ratio of Type 2 to Type 1
quasars is predicted to be between 1 and 2, more consistent with the
recent observational limits. The model also predicts that low-$L_X$ AGN
at high $z$
will almost always be completely obscured, a situation that would be
difficult to obtain for a standard unevolving circumnuclear AGN torus.

\subsubsection{The \citet{rms99} \nh\ Distribution}
\label{subsub:rms}
In a survey of local Seyfert 2 galaxies, \citet{rms99} found that
$\sim 75$\% had $\log N_{\mathrm{H}} \geq 23$, and about half were
Compton thick. We computed CXRB models using the same values of
$\alpha$ and $\beta$ as in the previous section, but now assuming the
\citet{rms99} \nh\ distribution for Type 2 objects (the \nh\
distribution for Type~1 AGN was unchanged). This results in 50\% more
Compton thick objects than with the previous flat distribution. 

\begin{figure}
\epsscale{1.0}
\plottwo{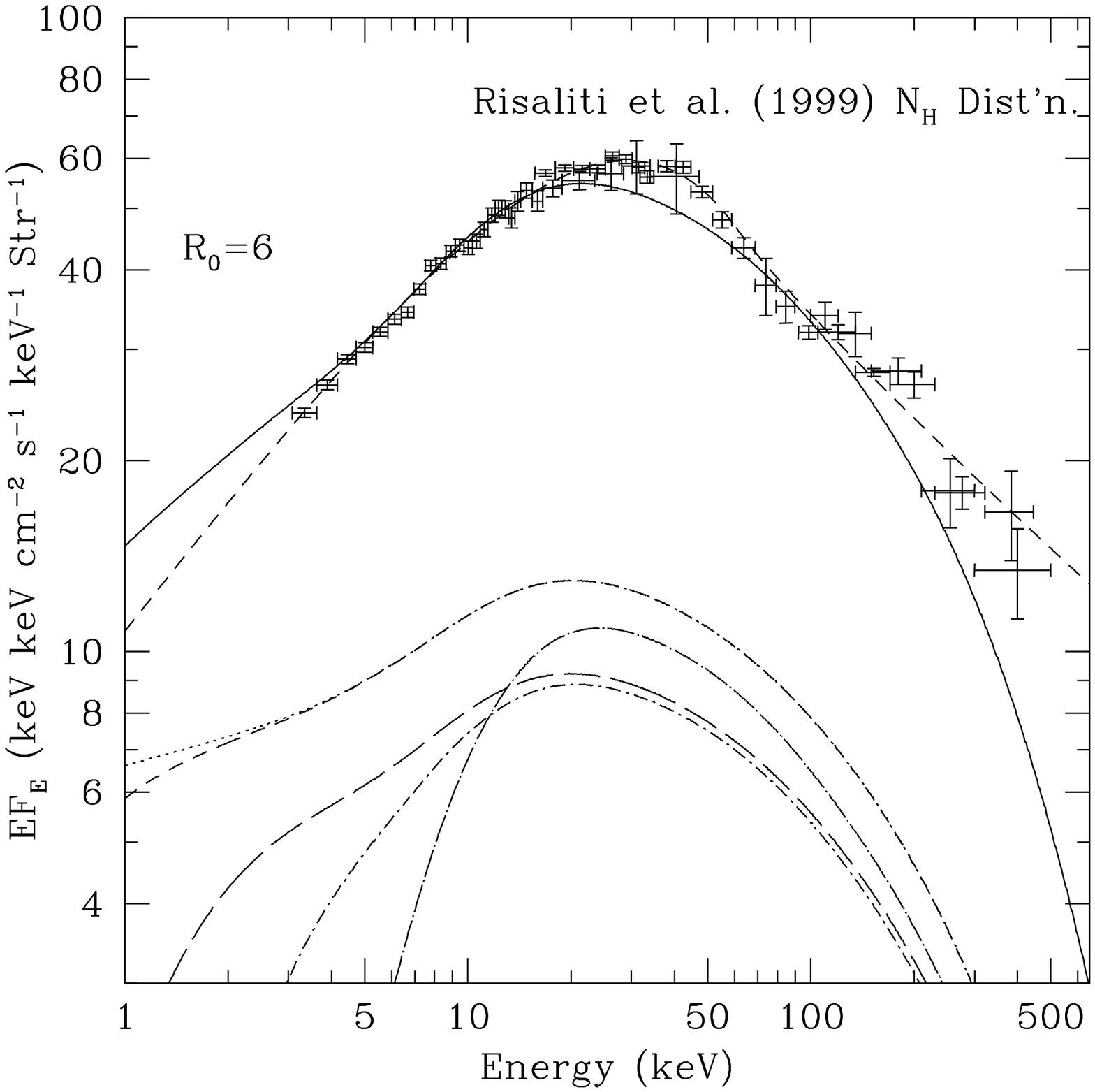}{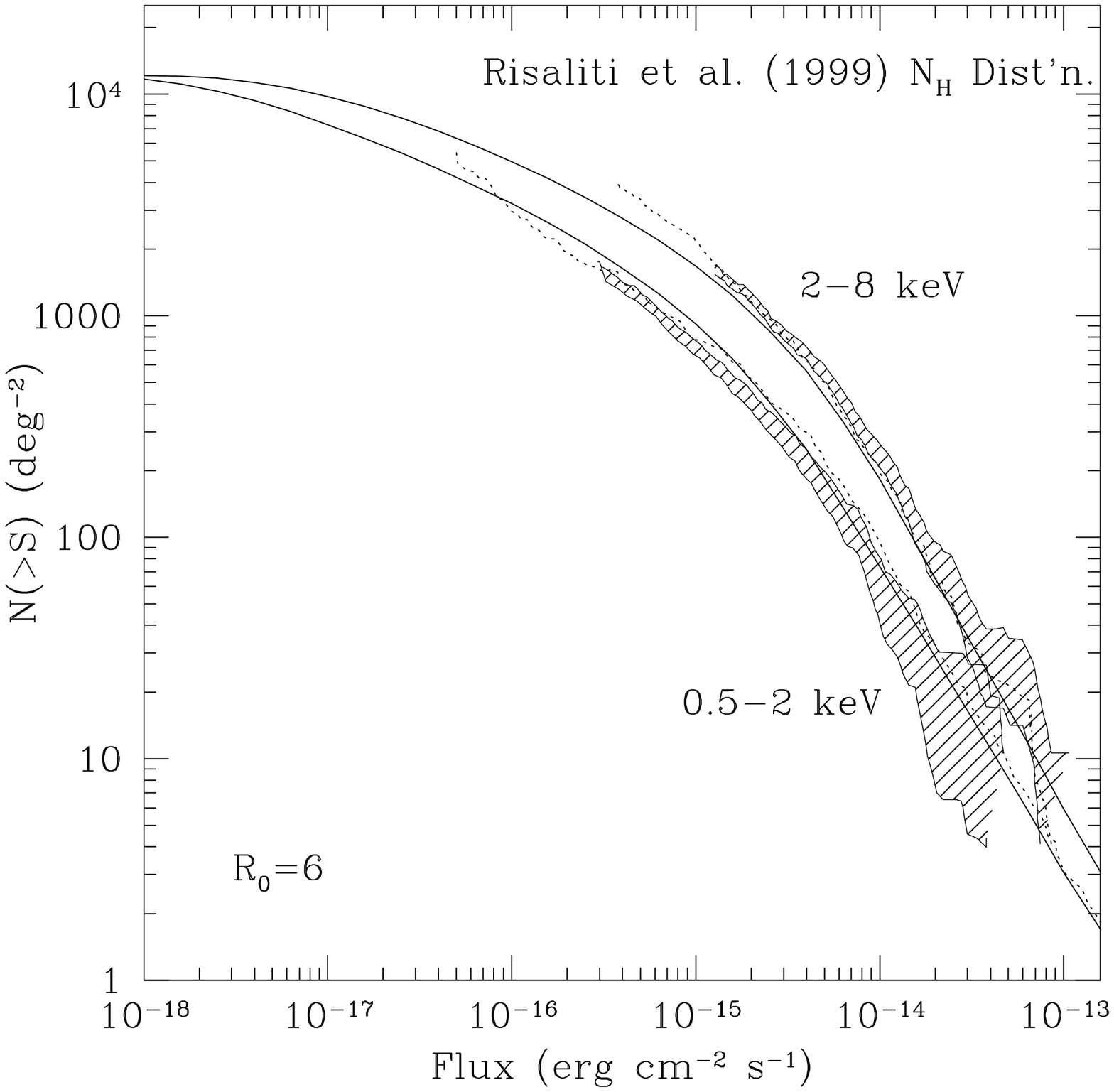}
\caption{Results from the power-law parameterization (Eq.~\ref{eq:plawmodel})
  when $R_0=6$ (as in Fig.~\ref{fig:6to1}) except the \citet{rms99}
  \nh\ distribution is assumed. The model spectra were multiplied by a
  factor of 1.57.}
\label{fig:6to1risaliti}
\end{figure}

The photon-index between 1 and 10~\kev\ for the predicted CXRB spectra
for these models, denoted $\Gamma_{1-10}^{\mathrm{Risaliti}}$, is
listed in the final column of Table~\ref{table:constantalpha}. In
general, there are only slight differences in the spectral slopes from
the two \nh\ distributions. For those cases when the Barger data is
fit very well and Type 2 AGNs are not dominant then the spectral slope
is slightly softer than before. This is because more absorbed AGN are
Compton thick and thus have little effect on the spectral shape below
10~\kev\ (see Fig.~\ref{fig:6to1risaliti}). However, if Type 2 AGNs are
the dominant population, as in cases which fit the Barger data poorly,
the additional Compton thick sources do slightly harden the
1--10~\kev\ spectral slope.

\begin{figure}
\epsscale{1.0}
\plottwo{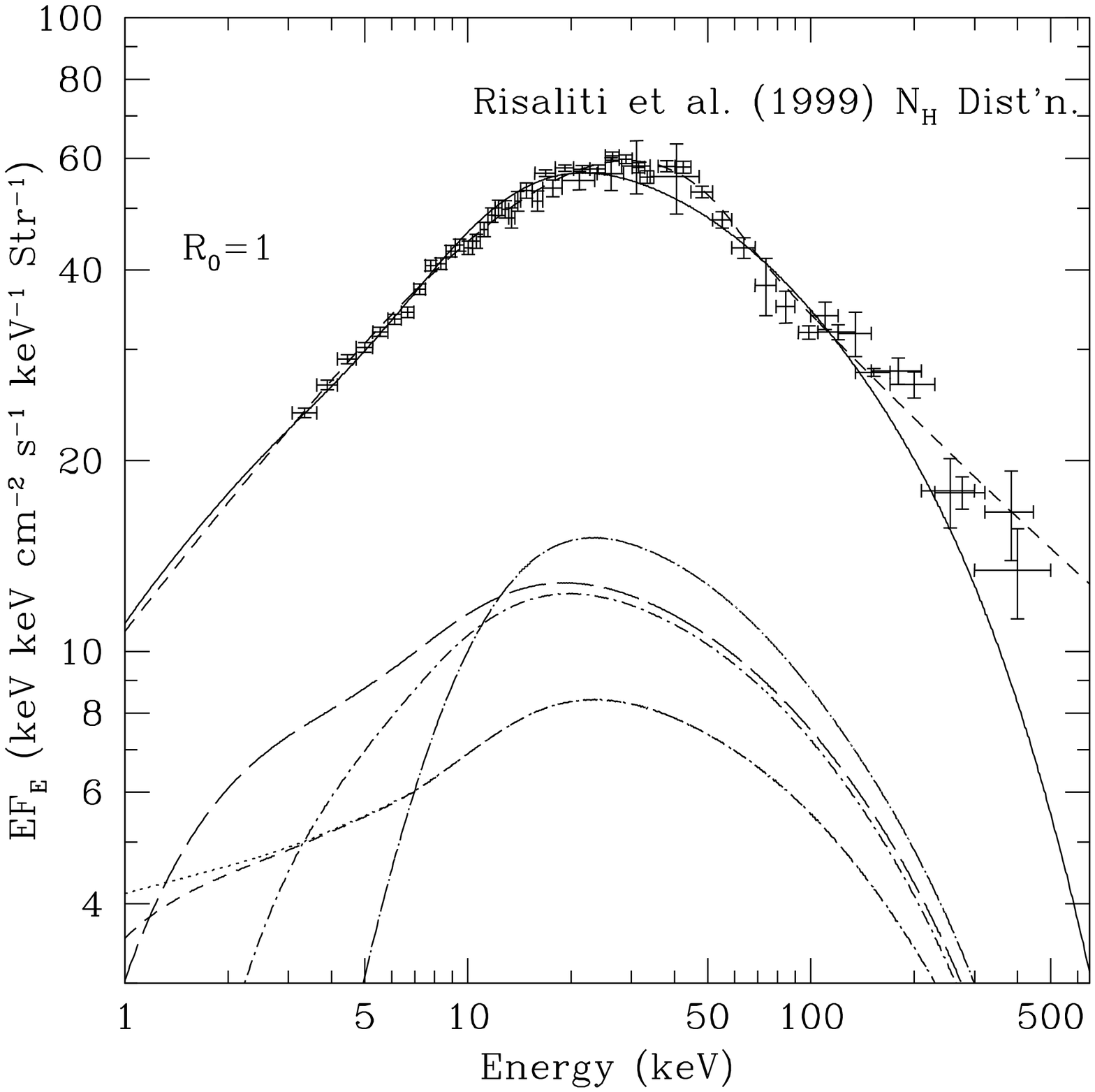}{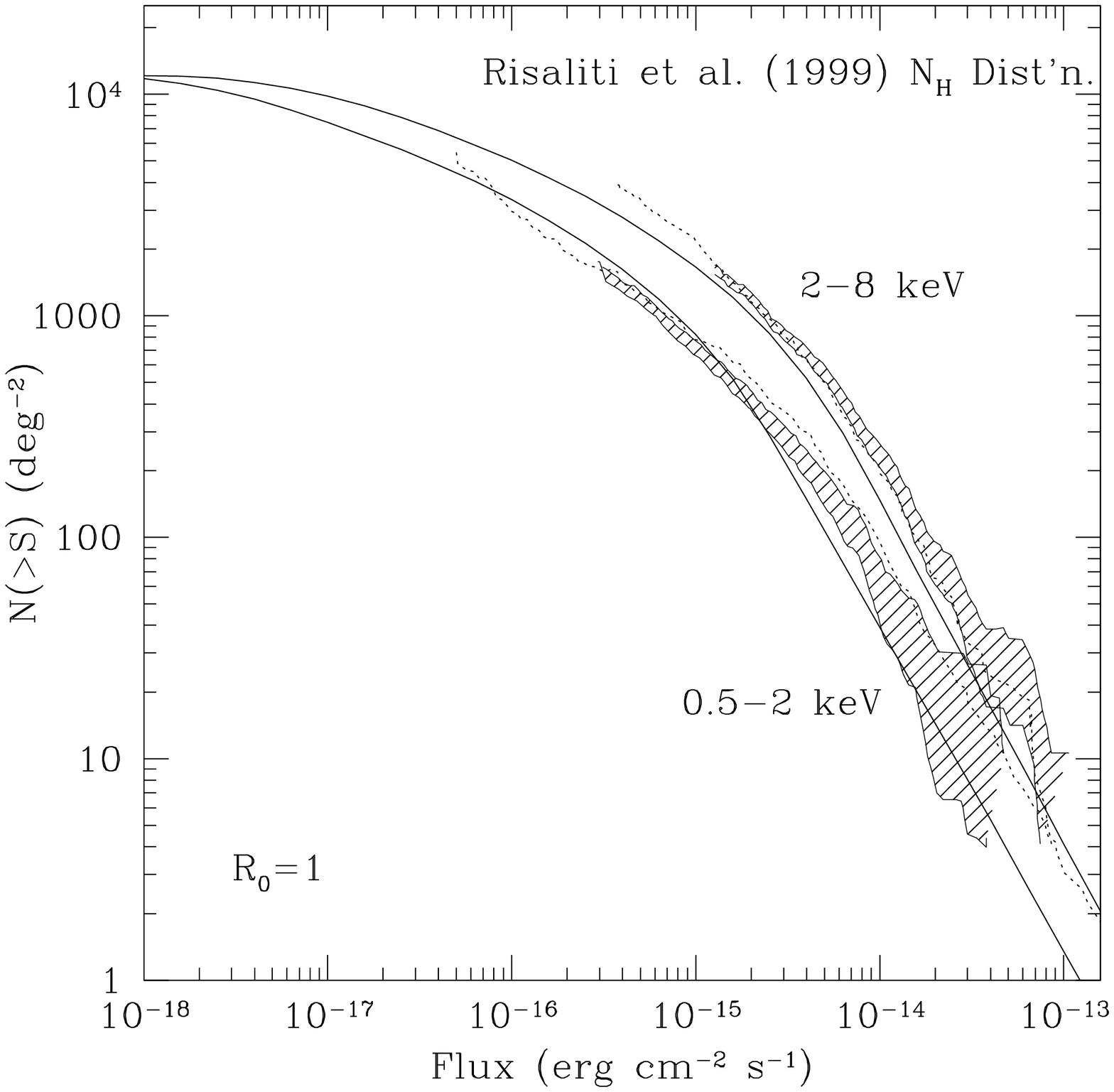}
\caption{Results from the $\log L_X$ power-law parameterization (Eq.~\ref{eq:loglmodel})
  when $R_0=1$ (as in Fig.~\ref{fig:1to1}) except the \citet{rms99}
  \nh\ distribution is assumed. The model spectra were multiplied by a
  factor of 1.7.}
\label{fig:1to1risaliti}
\end{figure}

\begin{figure}
\epsscale{1.0}
\plottwo{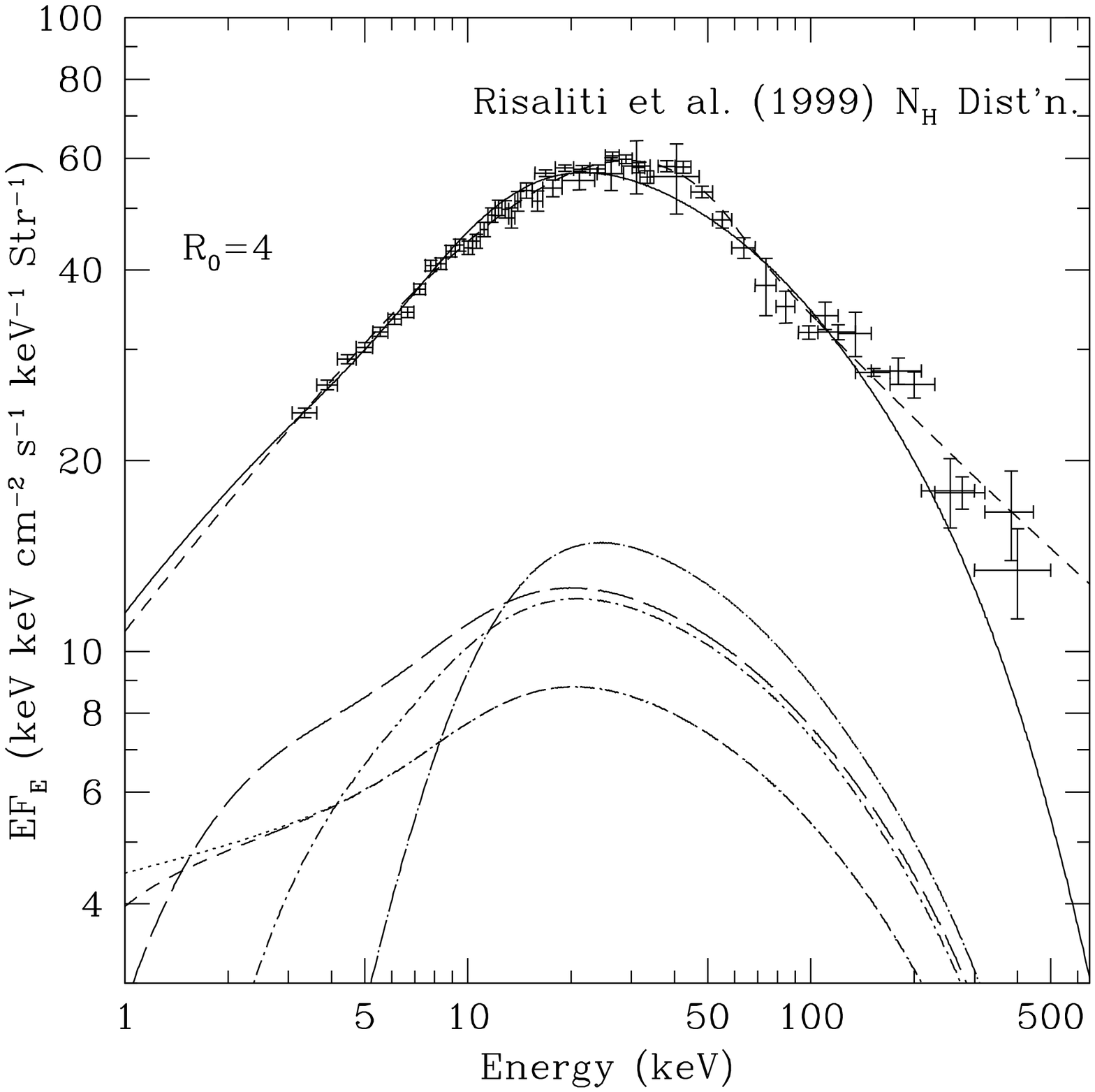}{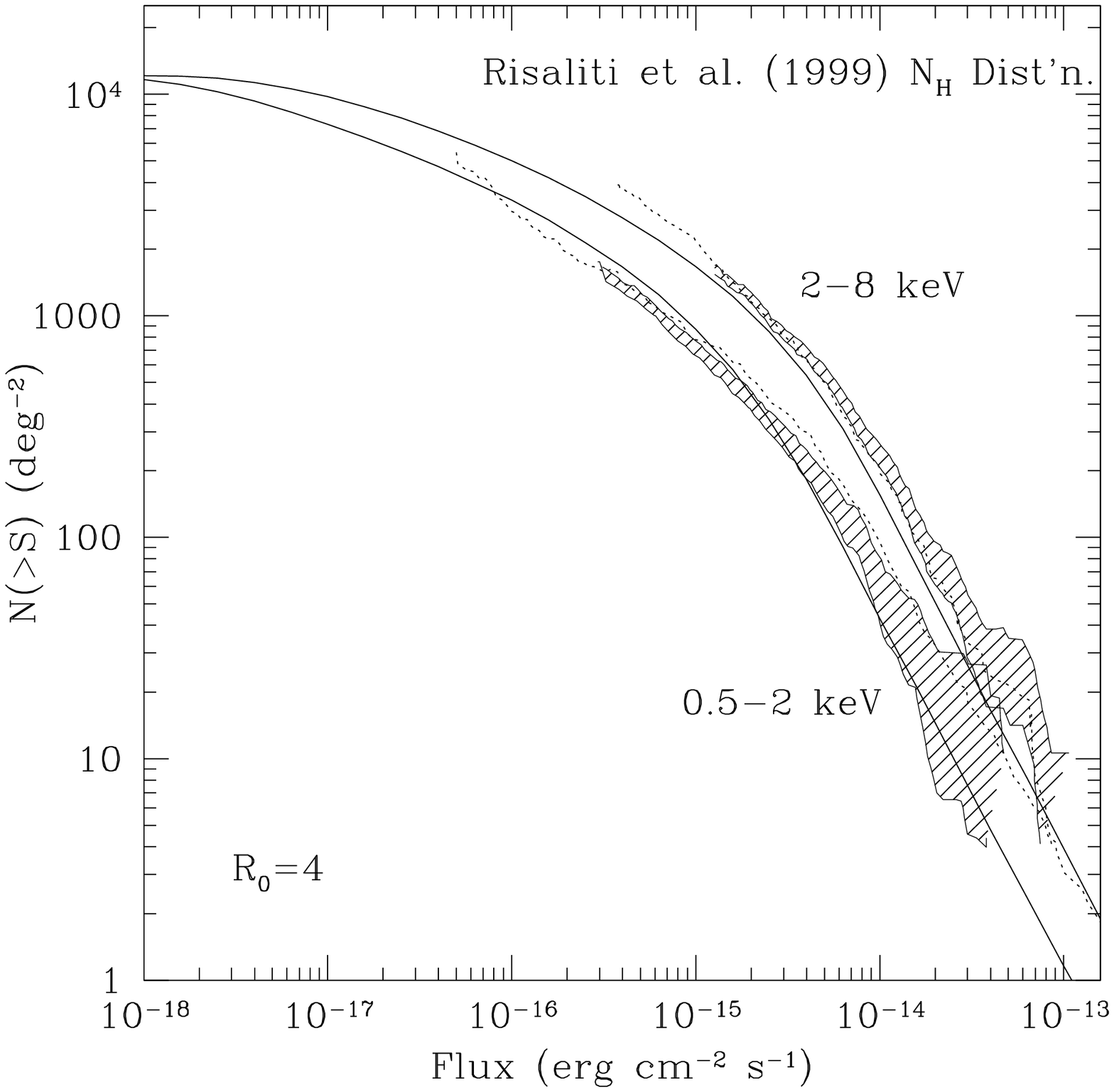}
\caption{Results from the $\log L_X$ power-law parameterization (Eq.~\ref{eq:loglmodel})
  when $R_0=4$ (as in Fig.~\ref{fig:4to1}) except the \citet{rms99}
  \nh\ distribution is assumed. The model spectra were multiplied by a
  factor of 1.7.}
\label{fig:4to1risaliti}
\end{figure}

While the effect on spectral shape below 10~\kev\ may be minimal, the
additional Compton thick AGN will significantly influence the
predicted spectrum at higher energies, particularly when Type 2
objects are dominant. This is illustrated in
Figures~\ref{fig:1to1risaliti} and~\ref{fig:4to1risaliti} where we
plot the predicted CXRB spectrum and number count distributions for
the $\log L_X$ power-law model when $R_0=1$ and $4$, respectively. These plots
are the equivalent of ones from Figs.~\ref{fig:1to1}
and~\ref{fig:4to1}, but with the \citet{rms99} \nh\ distribution. One
can clearly see that the large number of Compton thick sources in the
Risaliti distribution
provides a much better fit to the peak of the observed CXRB spectrum
at 40--50~\kev\ than the simple \nh\ model used previously. However,
the additional absorption also impacts the number counts,
and we find that in both the $R_0=1$ and $4$ models, the 2--8~\kev\
number counts underpredict the observations at all fluxes. These
results imply that additional Compton thick sources above the simple
\nh\ model are necessary to best fit the CXRB spectrum, but that the
\citet{rms99} distribution cannot hold at all $z$ and $L_X$. 

\subsection{Is $z$ evolution required?}
\label{sub:noz}
The above results show that we have managed to account
for the \citet{bar05} data, the CXRB spectrum, and the X-ray number counts
with a model that assumes $z$ evolution of $R$. However, it is not
clear if the data require such evolution; indeed, successful fits of
the CXRB spectrum and number counts have been obtained with no $z$
evolution of $R$ \citep{gf03,tu05}. The Barger data provides an additional
constraint, so to test if $z$ evolution is needed to fit these data,
we removed the $z$-dependence from
equations~\ref{eq:expmodel}-\ref{eq:cosmodel} and performed new
$\chi^2$ fitting for the luminosity parameter $\beta$. The results are
shown in Table~\ref{table:nozdependence} and should be compared with the
results in Table~\ref{table:constantalpha}. In every case but one, the
fit with no redshift dependence is worse, often by a large amount,
than the corresponding one with $z$ dependence. In fact, there are
only 18 models with no $z$-dependence that resulted in $P_{\chi^2} > 0.1$, as
compared to 33 when redshift evolution was allowed. In addition, when adequate
fits were obtained they required $R_0 > 6$, which reduces their
plausibility. Our conclusion is that, to the extent that the
\citet{bar05} data is an accurate depiction of $R$ from $z=0-1$, then
the data is best described by a model that includes redshift evolution
of the Type 2/Type 1 AGN ratio. Moreover, if the Sloan data are
accurate and $R_0 \approx 1$ then strong redshift evolution is required to
fit the CXRB.

\subsection{Luminosity-dependent $z$ evolution}
\label{sub:variablealpha}
The LDDE models of the X-ray AGN luminosity function
\citep{ueda03,has05} indicate that the most massive black holes,
typically responsible for luminous quasars, form earlier than less
massive black holes which power Seyfert galaxies (i.e.,
anti-hierarchal formation; \citealt{mer04,tou05}). Also,
radiative and wind-driven feedback will be more important for luminous
AGNs and will combat the tendency for increasing obscuration. Thus, it may be possible
that the redshift evolution of the Type 2/Type 1 AGN ratio may also
depend on luminosity. To test this, we allow the $\alpha$ parameter in
our evolutionary models (Eqs.~\ref{eq:expmodel}-\ref{eq:cosmodel}) to
be dependent on $\log L_X$ via a simple linear relation:
$\alpha(L_X)=m\log L_X + b$. The slope and intercept of this line was
determined by fixing $\alpha$ at $\log L_X=41.5$, denoted
$\alpha_{41.5}$, to the appropriate value in
Table~\ref{table:constantalpha}, and fit the Barger data for
$\alpha_{43.5}$, the value of $\alpha$ at $\log L_X=43.5$, along with
$\beta$. The results of this exercise are shown in
Table~\ref{table:variablealpha}. In all cases, the quality of the fits
are equal to or better than the ones where $\alpha$ was independent of
$L_X$, although the two models are statistically identical.

In terms of the direction of the $L_X$-dependence, the results are
inconclusive. The exponential evolutionary model prefers a positive
slope (i.e., the covering factor increases faster with $z$ at higher
$L_X$) when $R_0 < 5$, and zero slope in all other cases. The power-law
model gives a negative slope
(slower increase of the covering factor with $z$ at higher $L_X$) for
$R_0 < 8$ and a positive one for $R_0 \geq 8$. The cosine model seems
to vary from zero slope to slightly positive, but the $\log L_X$
power-law model consistently produces a negative slope, often with
$\alpha_{43.5}=0.0$, which implies \emph{negative} evolution of the
covering factor with $z$ for large $L_X$ AGN. Our conclusion is that the
available data is not constraining enough to presently determine if there is a
luminosity dependence to the redshift evolution of $R$. There is
tentative evidence that a negative slope is preferred, which, as
mentioned above, indicates that obscuration increases more slowly with
$z$ for more luminous AGNs. This can be understood if the
star-formation rate in these more luminous AGNs does not increase very
rapidly from $z=0-1$. 

\section{Discussion}
\label{sect:discuss}
In the previous section, we found that an
evolving Type 2/Type 1 AGN ratio, $R(L_X,z)$, determined by fitting the
Type 1 ratios observed by \citet{bar05} can account for the CXRB
spectrum and number counts.  Indeed, models with no $z$ dependence
provide significantly worse fits to Barger's data. Here, we discuss the
impact of this result on the observations, AGN unification and evolution, as well
as the connection to galaxy evolution.

\subsection{Implications for X-ray and Optical Observations}
\label{sub:observations}

Section~\ref{sect:results} presented moderately strong evidence for redshift
evolution of the obscuring medium around an AGN, but the strength of
that evolution depends on the assumption on what the local Type 2/Type
1 ratio is at low $L_X$. On one extreme, if we accept the Sloan results
that $R_0=1$ \citep{hao05,heck05}, then very strong $z$ evolution is
needed ($f_2 \propto (1+z)^{0.9}$) in order to account for the CXRB
spectrum (Figs.~\ref{fig:1to1} and~\ref{fig:1to1risaliti}). However,
this parameterization completely fails to fit the Type 1 ratios observed by
\citet{bar05}, requiring the X-ray surveys to miss many low-luminosity
unobscured sources and high luminosity obscured AGN. On the other
hand, models which best fit the Barger data (e.g.,
Figs.~\ref{fig:6to1} and~\ref{fig:6to1risaliti}) cannot reproduce the
shape of the CXRB spectrum for any $R_0$ --- there are too many
unobscured AGN.

We conclude that the best compromise solution is the $R_0=4$ model
shown in Figs.~\ref{fig:4to1} and~\ref{fig:4to1risaliti}. This parameterization
provides an adequate fit to the Barger data and a good match to the
observed shape of the CXRB spectrum and X-ray number counts. It
predicts a redshift evolution of the Type 2 AGN fraction of $f_2
\propto (1+z)^{0.3}$. We prefer this solution over the $R_0=1$
Sloan-inspired model because the value of $R_0$ is consistent with
earlier work \citep{mr95} and it requires fewer missing obscured
sources at higher $z$. The implication is that the Sloan work is
missing local obscured sources. These could be AGN whose narrow-line
regions are obscured and reddened, and therefore only appear as AGN in
the X-ray (i.e., X-ray Bright Optically Inactive Galaxies or elusive
AGN; \citealt{mai03,com02,gg05}). Future observational work is
required to accurate measure the local value of $R$ over a range in
$L_X$. This type of measurement is vital to the understanding of AGN and
galaxy evolution -- only when the local Type 2/Type 1 ratio is well
known can the amount of evolution be determined.

As described by \citet{bh05}, it is likely that the deep X-ray surveys
are missing heavily obscured AGN, especially ones that are Compton
thick. Evidence for this can be seen in the hard X-ray number counts,
which have yet to exhibit a turnover \citep{ros02}, and the stacking analysis
of \citet{wor05}, who shows that the CXRB may only be $\sim 50$~\% resolved
at $\sim 10$~\kev. We propose that the missing obscured AGN predicted
by the evolution model will mostly be Compton thick, but that the Type
2/Type 1 ratio will only be $\sim 1-2$ at higher AGN luminosities
(Fig.~\ref{fig:contour}). 

Comparing the results of the CXRB model between a simple uniform \nh\
distribution and the \citet{rms99} one, where half of the Seyfert 2s
are Compton thick, showed that additional Compton thick sources over
the uniform model are needed to improve the fit to the CXRB spectrum
at $\sim 40-50$~\kev. However, the \citet{rms99} ratio is too large to
hold in general, because the 2--8~\kev\ number counts are
significantly underpredicted. Similar to $R$, the fraction of
Compton-thick Seyfert 2s is likely to be a function of $z$ and
$L_X$. Thus, measurements of the \nh\ distribution at different
redshifts and luminosities are needed to accurately constrain the
predictions of CXRB synthesis models. A
focusing hard X-ray instrument, such as
\textit{NuStar}\footnote{http://www.nustar.caltech.edu/}, is necessary to
discover and trace the evolution of the Compton thick AGN population.

\subsection{AGN Unification, AGN Evolution and Galaxy Evolution}
\label{sub:unification}
These three topics are discussed together because our results, when
combined with recent observational and theoretical work, now show that
they are connected and dependent processes. One cannot discuss the
formation and evolution of a galaxy without accounting for the growth of
its black hole \citep{kh00}. Similarly, we assert that the evolution
of AGN and their observational appearance is connected to the evolution
of the host galaxy.

The fact that the ratio of obscured to unobscured AGN increases with
redshift requires a change to the traditional unification
paradigm\footnote{However, the basic axiom of the unification model
(namely, that all AGN are basically identical at the level of the
central engine) is unchanged by this result; only the primary source
of the obscuration needs to be altered.}. Since the obscuring medium
is changing with $z$ it must be influenced by forces and evolution on
the extragalactic scale. The cosmic star formation rate, which peaks
at a very similar redshift as $R$, seems to be the most likely
candidate. As star formation increases in a galaxy it signals an
increase in the
amount of absorbing gas and dust present that can act as an obscuring
medium for any AGN. The location of the absorbing material would
depend on the location and size of the star forming regions and could
exist close to the dust sublimation radius (and thus mimic some
properties of an absorbing torus), but could also be spread over a
large extent in the inner part of the galaxy \citep{mcleod95}. The idea of an extended,
more galactic-scale obscuring medium is consistent with recent
infra-red (IR) observations which have pointed out the remarkable
similarity in the mid-to-far-IR emission between Seyfert 2s and 1s
\citep{kur03,lu04}, in contrast to the predictions of the simple
molecular torus model. Of course, galactic scale obscuration of AGN
emission is commonly seen in the X-ray band when observing the
massively-star-forming Ultra-Luminous InfraRed Galaxies (ULIRGs; e.g.,
\citealt{iwa05}).

If, as seems to be the case, galactic star formation and AGN activity
are related then the question of AGN fueling arises (see recent review
by \citealt{jogee05}). At this point it is important to distinguish
between high luminosity quasars and lower luminosity Seyferts, which
dominate the production of the CXRB. As is well known, quasars peak in
density at high redshift ($z \sim 2$), and here interactions between
gas-rich disk galaxies are more common \citep{con03} and would cause
the ignition of a substantial starburst and accretion onto a black
hole \citep[e.g.,][]{hop05}. Indeed, only these most luminous AGNs
show observational evidence for interactions
\citep{hut87,dis95,bah97,kir99}. At lower luminosities there is little
observational evidence for Seyfert galaxies being associated with
mergers \citep{ls95,sch01,grog05}. Likewise, there is no evidence that
the decline in star formation from $z \sim 0.7$ is due to a decrease
in major mergers \citep{bell05}. Rather, it is possible that at lower
redshifts most interactions of a large disk galaxy are with smaller,
gas poor satellites \citep{mh94,wolf05,was05} that may trigger an
accretion phase onto the black hole, but with a much lower amount of
obscuration regulated by the associated star
formation. Therefore, it is not valid to apply the results of major
merger calculations to observations like the CXRB which are dominated
by Seyferts.

Depending on the exact nature of the interaction, a number of
different obscuration geometries could be set up around the black
hole, from a nearby starburst disk to an extended gas/dust lane
\citep[cf.,][]{bwm03}. Also, because
fueling a black hole is difficult ('the angular momentum problem';
\citealt{jogee05}), a very long time may pass following the
interaction before any gas makes it close enough to the center to be
accreted rapidly onto the hole. This delay between the causal event
and the resulting AGN activity could mask the cause from
observations, particularly if the interaction is weak. Thus, the
galaxy may not look like it was disrupted by a merger or interaction,
but the resulting star formation and AGN activity could still be
ongoing. 

The power of the CXRB spectrum and the deep
X-ray surveys is that they allow an investigation of the global properties
of AGN over cosmic time. We now know that most of the accretion in
the Universe is obscured, and that this obscuration evolves similar to
the star-formation rate. These facts deepen
the connection between star-formation and AGN fueling as well as
between black hole growth and galaxy evolution. To fully explore and
understand these connections, the sources of the CXRB must be found and
investigated in more detail over a range of redshift and luminosity.

\section{Conclusions}
\label{sect:concl}
This paper explored the hypothesis that the obscuration around an AGN
evolves with both redshift and luminosity. The motivation behind this
idea is that the obscuration is due to star-formation
ongoing in the host galaxy. Our conclusions are the following:
\begin{itemize}
\item To the extent that the Type~1 AGN fractions observed by
  \citet{bar05} are accurate, redshift evolution of the Type~2/Type~1
  AGN ($R$) to $z \sim 1$ is required to best fit the data.

\item The evolutionary model that most successfully reproduced the Barger data, the CXRB
  spectrum, and the X-ray number counts had $R_0 \approx 4$,
  consistent with \citet{mr95}. The redshift evolution of the Type 2
  AGN fraction is $(1+z)^{0.3}$. 

\item These models imply that the deep X-ray surveys are missing about
  50\% of obscured AGN with $\log L_X > 44$ at all $z$. These are likely
  to be mostly Compton thick.

\item Additional Compton-thick sources above the uniform \nh\
  distribution improve the fit to the CXRB spectrum. However, applying the
  \citet{rms99} distribution, where 50\% of all Type 2s are Compton
  thick, over all $L_X$ and $z$ underpredicts the 2--8~\kev\ number
  counts. The fraction of Compton thick obscured objects should be
  less than 50\% but will likely vary with $z$ and $L_X$.

\item There are very tentative indications that the covering factor
  evolves more slowly with $z$ for more luminous AGN.

\item A simple, non-evolving torus cannot alone provide the AGN obscuration
  over all cosmic time, and the shape of the CXRB spectrum is due to
  obscuration correlated with star-formation within the evolving host galaxy.

\item Seyfert galaxies, which dominate the production of the CXRB, are
  likely fueled by minor mergers or interactions which trigger a star
  formation event somewhere in the nuclear region. Unlike quasars, there may be a
  significant delay between the interaction and the subsequent
  ignition of the AGN.
\end{itemize}

\acknowledgments

This research was supported by the Natural Sciences and Engineering
Research Council of Canada. The authors thank E. Treister and M. Urry for helpful
discussions.

\clearpage

\begin{deluxetable}{ccccccc}
\tablewidth{0pt}
\tablecaption{\label{table:constantalpha}Results Assuming Both $z$ and $L_X$ Evolution}
\tablehead{
\colhead{$R_0$} & \colhead{$\alpha$} & \colhead{$\beta$} &
\colhead{$\chi^2$} & \colhead{$P_{\chi^2}$} &
\colhead{$\Gamma_{1-10}$} & \colhead{$\Gamma_{1-10}^{\mathrm{Risaliti}}$}}
\startdata
\cutinhead{$f_1=K \exp \left (L_X \over {10^{\beta} (1+z)^{\alpha}} \right )$}
1 & 14.9 & 42.5 & 215 & $6\times 10^{-40}$ & 1.574 & 1.578 \\
2 & 12.2 & 42.5 & 70.9 & $8\times 10^{-11}$ & 1.505 & 1.505 \\
3 & 11.1 & 42.5 & 31.8 & $8\times 10^{-4}$ & 1.463 & 1.460 \\
4 & 5.4 & 43 & 18.4 & 0.072 & 1.501 & 1.504 \\
5 & 5.1 & 43 & 11.4 & 0.41 & 1.487 & 1.489 \\
6 & 4.9 & 43 & 9 & 0.62 & 1.474 & 1.476 \\
7 & 4.8 & 43 & 8.8 & 0.64 & 1.461 & 1.462 \\
8 & 5.1 & 42.9 & 9.5 & 0.58 & 1.453 & 1.452 \\
9 & 5 & 42.9 & 10.5 & 0.49 & 1.446 & 1.444 \\
10 & 4.9 & 42.9 & 11.7 & 0.39 & 1.440 & 1.438 \\
\cutinhead{$f_1=K(1+z)^{-\alpha} \left (L_X \over 10^{41.5} \right
  )^{\beta}$}
1 & 3.5 & 0.25 & 15.3 & 0.17 & 1.477 & 1.473 \\
2 & 3 & 0.3 & 11.6 & 0.39 & 1.489 & 1.488 \\
3 & 2.6 & 0.33 & 9.6 & 0.57 & 1.50 & 1.501 \\
4 & 2.3 & 0.36 & 8.3 & 0.69 & 1.516 & 1.519 \\
5 & 2.1 & 0.38 & 7.5 & 0.76 & 1.52 & 1.525 \\
6 & 1.9 & 0.39 & 7.0 & 0.8 & 1.52 & 1.525 \\
7 & 1.7 & 0.39 & 6.7 & 0.82 & 1.513 & 1.519 \\
8 & 1.5 & 0.4 & 6.5 & 0.84 & 1.521 & 1.528  \\
9 & 1.4 & 0.41 & 6.3 & 0.85 & 1.521 & 1.528 \\
10 & 1.3 & 0.42 & 6.3 & 0.85 & 1.523 & 1.530 \\
\cutinhead{$f_2=K(1+z)^{\alpha} (\log L_X)^{-\beta}$}
1 & 0.9 & 1.3 & 28.4 & $2.8\times 10^{-3}$ & 1.412 & 1.402 \\
2 & 0.5 & 2.7 & 18.4 & 0.072 & 1.407 & 1.398 \\
3 & 0.4 & 4.7 & 15.2 & 0.18 & 1.423 & 1.419 \\
4 & 0.3 & 4.8 & 13.8 & 0.24 & 1.418 & 1.413 \\
5 & 0.3 & 6.3 & 13.1 & 0.29 & 1.434 & 1.432 \\
6 & 0.2 & 5.1 & 12.7 & 0.32 & 1.415 & 1.411 \\
7 & 0.2 & 5.9 & 12.2 & 0.35 & 1.425 & 1.422 \\
8 & 0.2 & 6.5 & 11.9 & 0.37 & 1.431 & 1.429 \\
9 & 0.2 & 6.9 & 11.8 & 0.38 & 1.434 & 1.433 \\
10 & 0.2 & 7.3 & 11.7 & 0.39 & 1.439 & 1.438 \\
\cutinhead{$f_2=K(1+z)^{\alpha} \cos^2 \left( {\log L_X -41.5 \over
    \beta} \right )$}
1 & 0.9 & 5.5 & 25.0 & $9\times 10^{-3}$ & 1.467 & 1.465 \\
2 & 0.5 & 4.4 & 13.8 & 0.24 & 1.474 & 1.475 \\
3 & 0.3 & 4.4 & 11.2 & 0.43 & 1.466 & 1.468 \\
4 & 0.2 & 4.3 & 10.1 & 0.52 & 1.465 & 1.467 \\
5 & 0.2 & 3.8 & 8.9 & 0.63 & 1.480 & 1.484 \\
6 & 0.1 & 4.2 & 9.1 & 0.61 & 1.461 & 1.464 \\
7 & 0.1 & 3.9 & 8.3 & 0.69 & 1.472 & 1.476 \\
8 & 0.1 & 3.7 & 8.1 & 0.70 & 1.481 & 1.486 \\
9 & 0.1 & 3.6 & 8.2 & 0.69 & 1.484 & 1.490 \\
10 & 0.1 & 3.5 & 8.5 & 0.67 & 1.489 & 1.495\\ 
\enddata
\tablecomments{$P_{\chi^2}$ is the null-hypothesis probability for the
calculated $\chi^2$. $\Gamma_{1-10}$ is the photon-index between 1 and
10~\kev\ of the CXRB spectrum predicted by that particular
model. $\Gamma_{1-10}^{\mathrm{Risaliti}}$ is the photon-index when
the \citet{rms99} \nh\ distribution is used. }
\end{deluxetable}

\clearpage

\begin{deluxetable}{cccc}
\tablewidth{0pt}
\tablecaption{\label{table:nozdependence}Results Assuming only $L_X$ Evolution}
\tablehead{
\colhead{$R_0$} & \colhead{$\beta$} & \colhead{$\chi^2$} & \colhead{$P_{\chi^2}$} }
\startdata
\cutinhead{$f_1=K \exp \left (L_X \over {10^{\beta}} \right )$}
1 & 46.9 & 279 & $10^{-52}$ \\
2 & 44.8 & 96 & $4\times 10^{-15}$ \\
3 & 44.6 & 43.3 & $2\times 10^{-5}$ \\
4 & 44.6 & 24.8 & 0.016 \\
5 & 44.5 & 17.8 & 0.12 \\
6 & 44.5 & 15.8 & 0.20 \\
7 & 44.4 & 15.8 & 0.20 \\
8 & 44.4 & 16.4 & 0.18 \\
9 & 44.4 & 17.5 & 0.13  \\
10 & 44.4 & 19.0 & 0.088  \\
\cutinhead{$f_1=K \left (L_X \over 10^{41.5} \right )^{\beta}$}
1 & 0.0 & 279 & $10^{-52}$ \\
2 & 0.0 & 97 & $2\times 10^{-15}$ \\
3 & 0.0 & 46.4 & $6\times 10^{-6}$ \\
4 & 0.0 & 29.1 & $3.8\times 10^{-3}$ \\
5 & 0.01 & 23.0 & 0.028 \\
6 & 0.08 & 19.6 & 0.076 \\
7 & 0.14 & 16.7 & 0.16 \\
8 & 0.18 & 14.4 & 0.27 \\
9 & 0.21 & 12.6 & 0.40 \\
10 & 0.24 & 11.2 & 0.51 \\
\cutinhead{$f_2=K (\log L_X)^{-\beta}$}
1 & 0.0 & 279 & $10^{-52}$ \\
2 & 0.0 & 97 & $2\times 10^{-15}$ \\
3 & 0.0 & 46.4 & $6\times 10^{-6}$ \\
4 & 0.0 & 29.1 & $3.8\times 10^{-3}$ \\
5 & 0.1 & 23.0 & 0.028 \\
6 & 1.0 & 20.2 & 0.063 \\
7 & 1.7 & 18.4 & 0.10 \\
8 & 2.2 & 17.2 & 0.14 \\
9 & 2.7 & 16.2 & 0.18 \\
10 & 3.0 & 15.5 & 0.21 \\
\cutinhead{$f_2=K \cos^2 \left( {\log L_X -41.5 \over \beta} \right )$}
1 & 39.9 & 280 & $8\times 10^{-53}$ \\
2 & 39.9 & 97 & $2\times 10^{-15}$ \\
3 & 39.9 & 46.5 & $6\times 10^{-6}$ \\
4 & 9.7 & 28.6 & $4.5\times 10^{-3}$ \\
5 & 6.3 & 20.1 & 0.065 \\
6 & 5.3 & 15.5 & 0.22 \\
7 & 4.8 & 12.7 & 0.39 \\
8 & 4.5 & 11.0 & 0.53 \\
9 & 4.3 & 9.9 & 0.62 \\
10 & 4.1 & 9.2 & 0.69 \\ 
\enddata
\tablecomments{$P_{\chi^2}$ is the null-hypothesis probability for the
calculated $\chi^2$. Here, the number of degrees of freedom
is 12.}
\end{deluxetable}

\clearpage

\begin{deluxetable}{cccccc}
\tablewidth{0pt}
\tablecaption{\label{table:variablealpha}Results Assuming
  Luminosity-Dependent $z$ Evolution}
\tablehead{
\colhead{$R_0$} & \colhead{$\alpha_{41.5}$} &
\colhead{$\alpha_{43.5}$} & \colhead{$\beta$} &
\colhead{$\chi^2$} & \colhead{$P_{\chi^2}$}}
\startdata
\cutinhead{$f_1=K \exp \left (L_X \over {10^{\beta} (1+z)^{\alpha(L_X)}} \right )$}
1 & 14.9 & 17.9 & 42 & 120 & $1.7\times 10^{-20}$ \\
2 & 12.2 & 16.1 & 42 & 38.7 & $5.9\times 10^{-5}$ \\
3 & 11.1 & 15.2 & 42 & 19.5 & 0.052 \\
4 & 5.4 & 14 & 42 & 14.5 & 0.21  \\
5 & 5.1 & 5.1 & 43 & 11.4 & 0.41 \\
6 & 4.9 & 4.9 & 43 & 9 & 0.62 \\
7 & 4.8 & 4.8 & 43 & 8.8 & 0.64 \\
8 & 5.1 & 5.1 & 42.9 & 9.5 & 0.58 \\
9 & 5 & 5 & 42.9 & 10.5 & 0.49 \\
10 & 4.9 & 4.9 & 42.9 & 11.7 & 0.39 \\
\cutinhead{$f_1=K(1+z)^{-\alpha(L_X)} \left (L_X \over 10^{41.5} \right
  )^{\beta}$}
1 & 3.5 & 1.0 & 0.0 & 10.6 & 0.48 \\
2 & 3 & 1.0 & 0.09 & 9.0 & 0.63 \\
3 & 2.6 & 1.2 & 0.18 & 8.1 & 0.70 \\
4 & 2.3 & 1.3 & 0.24 & 7.6 & 0.75 \\
5 & 2.1 & 1.5 & 0.31 & 7.1 & 0.79 \\
6 & 1.9 & 1.5 & 0.34 & 6.8 & 0.81 \\
7 & 1.7 & 1.5 & 0.37 & 6.6 & 0.83 \\
8 & 1.5 & 1.6 & 0.41 & 6.5 & 0.84 \\
9 & 1.4 & 1.5 & 0.42 & 6.3 & 0.85 \\
10 & 1.3 & 1.6 & 0.45 & 6.2 & 0.86 \\
\cutinhead{$f_2=K(1+z)^{\alpha(L_X)} (\log L_X)^{-\beta}$}
1 & 0.9 & 0.8 & 0.0 & 2.5 & $4.7\times 10^{-3}$ \\
2 & 0.5 & 0.3 & 0.2 & 17.1 & 0.11 \\
3 & 0.4 & 0.0 & 0.0 & 13.5 & 0.26 \\
4 & 0.3 & 0.0 & 1.1 & 12.9 & 0.30 \\
5 & 0.3 & 0.0 & 2.6 & 11.9 & 0.37 \\
6 & 0.2 & 0.0 & 2.7 & 12.4 & 0.34 \\
7 & 0.2 & 0.0 & 3.4 & 11.8 & 0.38 \\
8 & 0.2 & 0.0 & 4.0 & 11.4 & 0.41 \\
9 & 0.2 & 0.0 & 4.5 & 11.2 & 0.43 \\
10 & 0.2 & 0.0 & 4.8 & 11.1 & 0.44 \\
\cutinhead{$f_2=K(1+z)^{\alpha(L_X)} \cos^2 \left( {\log L_X -41.5 \over
    \beta} \right )$}
1 & 0.9 & 0.9 & 5.5 & 25.0 & $9\times 10^{-3}$ \\
2 & 0.5 & 0.5 & 4.4 & 13.8 & 0.24 \\
3 & 0.3 & 0.4 & 3.9 & 10.5 & 0.49 \\
4 & 0.2 & 0.4 & 3.4 & 9.1 & 0.61 \\
5 & 0.2 & 0.2 & 3.8 & 8.9 & 0.63 \\
6 & 0.1 & 0.3 & 3.3 & 8.0 & 0.71 \\
7 & 0.1 & 0.2 & 3.5 & 8.0 & 0.71 \\
8 & 0.1 & 0.1 & 3.7 & 8.1 & 0.70 \\
9 & 0.1 & 0.1 & 3.6 & 8.2 & 0.69 \\
10 & 0.1 & 0.0 & 3.9 & 8.3 & 0.69 \\ 
\enddata
\tablecomments{$P_{\chi^2}$ is the null-hypothesis probability for the
calculated $\chi^2$. $\alpha_{41.5}$ is the value of $\alpha(L_X)$ at $\log
L_X =41.5$. $\alpha_{43.5}$ is the value of $\alpha(L_X)$ at $\log
L_X =43.5$}
\end{deluxetable}

\clearpage

\clearpage

\clearpage

\clearpage

\clearpage

\clearpage

\clearpage

\clearpage

\end{document}